\documentclass[12pt,preprint]{aastex}
\usepackage[english]{babel}
\usepackage[utf8x]{inputenc}
\usepackage[T1]{fontenc}

\usepackage{amsmath,bm}
\usepackage{amssymb}
\usepackage{graphicx}
\usepackage[colorlinks=true, allcolors=blue]{hyperref}
\usepackage{booktabs}
\usepackage{ulem}
\usepackage{lineno}

\shorttitle{Runaway Growth of Planetesimals: Revisited}
\shortauthors{Darriba \& Haghighipour}

\begin{document}

\title{Runaway Growth of Planetesimals Revisited:\\
\normalsize Presenting Criteria Required for Realistic Modeling of Planetesimal Growth}

\author{Nader Haghighipour\altaffilmark{1,2,3}  and Luciano A. Darriba \altaffilmark{4}}

\altaffiltext{1}{Planetary Science Institute, 1700 East Fort Lowell, Tucson, AZ 85719, USA} 
\altaffiltext{2}{Institute for Astronomy, University of Hawaii-Manoa, 2680 Woodlawn Dr.,
Honolulu, HI 96822, USA}
\altaffiltext{3}{Institute for Advanced Planetary Astrophysics, 6770 Hawaii Kai Dr., Honolulu, HI, USA}
\email{nader@psi.edu}
\altaffiltext{4}{Facultad de Ciencias Astron\'omicas y Geof\'\i sicas, Universidad Nacional de La Plata, 
Paseo del Bosque s/n (1900), La Plata, Argentina.}

\begin{abstract}
We have initiated a large project on identifying the requirements for 
developing a realistic and ground-up approach to simulating the formation of terrestrial planets in our solar system. 
As the first phase of this project, we present here the criteria that any model of planetesimal growth needs to fulfill 
in order to be self-consistent and produce reliable results. We demonstrate how these criteria emerge by revisiting 
runaway growth and carrying out a thorough analysis of its results. As our goal is to identify the pathway to a realistic model, 
we focus analysis on simulations where at the beginning, planetesimals are not artificially enlarged. We show how using 
uninflated planetesimals, as the first requirement for a realistic model, will result in a set of criteria naturally emerging 
from the evolution of the system. For instance, the growth times in simulations with uninflated planetesimals become comparable 
to the time of giant planet formation implying that any realistic simulation of planetesimal growth, in addition to using 
real-size planetesimals, needs to include the perturbation of the growing giant planets as well. Our analysis also points 
to a strong connection between the initial distribution of planetesimals and the final outcome. For instance,
due to their natural expansion, initially isolated distributions, or a collection of initially isolated distributions, 
such as rings of planetesimals, do not produce reliable results. In a self-consistent and realistic model, where the 
initial conditions are supported by basic principles and do not include simplifying, ad hoc assumptions, the entire disk of 
planetesimals has to be simulated at once. We present the results of our analyses and discuss their implied criteria. 
\end{abstract}

\section{Rationale}
\label{section:rationale}

Recent advances in computational technology have had major impacts on the studies of the formation
and evolution of planetary systems. Fast and powerful computers have enabled
high-resolution simulations, and sophisticated codes have allowed collisions to be modeled more accurately 
\citep[e.g.,][]{Burger20,Reinhardt22}.
Collectively, these developments have elevated the field to a new level where now the majority of the
underlying assumptions that have primarily been developed to circumvent computational complexities can be removed, 
and almost any physical processes relevant to planet formation can be directly included in the simulations. 
The latter is a powerful capability that allows for examining the consequences of different physical effects 
thereby developing a pathway to building a realistic formation model from grounds up. On that front, it is of 
utmost importance to determine the requirements that any model of planet formation needs to satisfy in order 
for its results to be realistic.

In the past few years, the idea of removing simplifying assumptions, combined with requiring models to meet criteria 
for producing realistic outcomes raised a demand from some members of the community to justify its worth and significance. It
is important to note that any action to respond to this demand needs to not only identify the requirements for a realistic model, 
but also to demonstrate, for each criterion, the scientific rational that would justify its consideration. To do so, the 
foundations of planet formation need to be revisited, and all their initial conditions and underlying assumptions need 
to be examined. The goal of this article is to address the above in the context of planetesimals runaway growth.

One of the most outstanding assumptions, that despite its unrealistic nature, has been used repeatedly for almost three
decades, is the assumption of ``inflated planetesimals.'' As we will explain in the next section, to avoid long integration 
times associated with the collisional growth of small bodies, in almost all simulations of planetesimal growth, the rate of 
the collisions of planetesimals is artificially enhanced by increasing their initial sizes, forcing their growth to happen 
in shorter times. Table 1 shows an up-to-date list of the studies that have used inflated planetesimals along with their 
regions of simulation, integration resolution (number of planetesimals), and their radius-enlargement factor $(f)$.
While this unphysical approach does not affect the mechanics of the occurrence of runaway growth, and may 
even be useful in demonstrating some of the underlying physics of the process, it will have profound effects on the formation 
of planetary embryos and the subsequent evolution of the system. For instance, in our solar system, it is the conventional 
wisdom that terrestrial planets formed through a two-stage process: the runaway growth, during which larger planetesimals swept 
up smaller ones and grew to Moon- to Mars-sized planetary embryos, followed by giant impacts among the latter 
bodies\footnote{It has been shown by \citet{Levison15a,Levison15b}, \citet{Matsumura17}, \citet{Lambrechts19}
and \citet{Johansen21} that terrestrial planets can also form via the Pebble Accretion scenario. However, it has 
also been shown by \citet{Morbidelli25} that this sceanrio does not satisfy the compositional, dynamical, and 
chronological constraints associated with the terrestrial planets of our solar system.}. Because runaway growth is a 
natural consequence of the dynamical evolution of planetesimals, when studying the formation of terrestrial 
planets, it is customary to, {\textit{a priori}}, assume its outcome and use the latter as the initial condition for
simulating the state of giant impacts. However, during this time (i.e., runaway growth), the dynamics of the system 
is strongly affected by the perturbation of its rapidly growing bodies. As the growth of these objects
has been artificially enhanced (by increasing the sizes of their seed planetesimals), their perturbation becomes 
stronger than it would have been, had their planetesimals not been inflated.  
The latter deviates the system from its natural evolution within its natural timescale, 
causing it to take a dynamical path that is dictated to it by the perturbation of its artificially hyper-sized bodies.
This deviation of the system from its natural evolution in the runaway growth phase propagates to the phase of giant impacts, 
rendering the outcome of the simulations of the late stage of terrestrial planet formation unreliable.
For instance, because the accretion zone of the artificially enlarged runaway growth bodies increases with their large masses, 
the planetary embryos that are formed by the growth of these objects become larger than they would have been without the radius 
inflation. This larger masses of planetary embryos profoundly affects the masses and orbital architecture of the final 
terrestrial planets as it increases the spacing among the embryos (an important factor in assembling protoplanetary disks at 
the beginning of the giant-impacts simulations) and decreases the effect of dynamical friction by scattering 
a large number of smaller bodies out of the system.

To ensure that the final system is free from these biases, simulations have to be carried out using
uninflated planetesimals. The results of such simulations, combined with the mechanics of the process will have 
implications that point to and/or manifest themselves as requirements for obtaining realistic outcomes. To better understand 
this process, its worthiness, and the significance of its implications, it is necessary to develop a deep 
understanding of the history of modeling planetesimal growth and more importantly, the thought-process through which 
the field has progressed. In this article, we aim to achieve the latter by demonstrating how the requirements for a realistic 
model emerge organically from the natural evolution of the system. To do so, we begin by presenting an in-depth review of the 
history of planetesimal growth in the next section, and continue by carrying out new simulations and analyzing the implications 
of the results. Because, traditionally, planetesimal growth has been studied as an introduction 
to terrestrial planet formation, we follow the same approach and focus our study and simulations within the same context. 
However, our analyses are completely general and can be applied to the formation of the cores of the giant planets as well.

\section{Introduction}
\label{section:introduction}

That the collision among small planetesimals may result in the rapid formation of much larger bodies
was first noted by \citet{Greenberg78}. Using the particle-in-a-box approach, these authors found that in a swarm 
of km-sized planetesimals, a few locally large bodies may dominate the process and rapidly grow to a few hundred 
km in size. Prior to that study, \citet{Safronov69} had shown that planetesimal growth would proceed 
in an orderly manner meaning that large planetesimals gradually grow by sweeping up smaller ones while  maintaining 
their mass-ratios. Orderly growth was also reported in later studies by \citet{Nakagawa83}, \citet{Ohtsuki88} and 
\citet{Hayakawa89} who advanced the work of \citet{Safronov69} by including the effect of gas drag.

The viability of runaway growth was first examined by \citet{Wetherill89}. 
Because the underlying motivation behind their study was to understand 
the formation of Earth, these authors considered a swarm of planetesimals between 0.98 AU and 1.02 AU, and
following \citet{Safronov62,Safronov69}, distributed them according to the mass distribution function

\begin{equation}
{{dN(m)}\over {dm}} \propto {e^{-m/{m_0}}}\,,
\end{equation}

\noindent
(similar distribution function had also been used by Greenberg et al. 1978 and Nakagawa et al. 1983).
In this equation, $N(m)$ is the number density of planetesimals with mass $m$, and ${m_0}$ is the 
mean mass of the planetesimals distribution at the time $t=0$.
\citet{Wetherill89} showed that, under the conditions considered by \citet{Safronov69} and \citet{Nakagawa83} (that is,
when the mutual interactions between planetesimals are neglected), the results obtained by those authors were correct. 
However, when simulations include planetesimals interactions, the equipartition of energy \citep{Stewart80}, 
resulted from, for instance, dynamical friction, viscus stirring, the existence of large bodies, low-velocity 
gravitational focusing, gas drag, and fragmentation, will cause a few large bodies to dominate the growth and, 
in a relatively short time, reach sizes of the order of several hundreds to a few thousands times those of the 
initial planetesimals.
\citet{Wetherill89} also carried out simulations similar to those in \citet{Greenberg78}. They 
showed that unlike the results reported by those authors, their simulations show no signs and indications of 
runaway growth. \citet{Wetherill89} attributed the appearance of runaway growth in the simulations by
\citet{Greenberg78} to the inaccuracy of the numerical method used by those authors \citep{Wetherill90}.

In principle, gravitational focusing, combined with the equipartition of energy
is the main reason that runaway growth occurs \citep{Ohtsuki93,Kokubo96}. 
In a system of planetesimals with different sizes, the equipartition of energy manifests 
itself in the form of dynamical friction. That is, the interaction between a large planetesimal and a population of 
smaller ones reduces the relative velocity of the large body by lowering its eccentricity and inclination
\citep{Ida90,Ida92a,Ida92b,Ida93}. This low relative velocity enhances the collisional growth of the large 
planetesimal, causing it to dominate the growth process in its surrounding. 

Although, as a proof of concept, simulations by \citet{Wetherill89} successfully demonstrated the appearance 
and progression of runaway growth, their underlying assumptions 
limited the applicability and generalization of their results. For instance,
these authors did not include the effect of the gravity of the Sun on the collision of 
objects and assumed, {\it a priori}, that the spatial velocities of planetesimals would follow a triaxial 
Gaussian distribution. These limitations motivated other researchers to revisit runaway growth and carry out
new simulations \citep{Ida90,Ohtsuki90,Barge91,Spaute91,Kolvoord92,Wetherill93}.
However, all these simulations, as well as those of \citet{Wetherill89} still suffered from one fundamental 
inconsistency: In all these studies, growth was modeled through a statistical approach 
using the coagulation equation in the context of kinetic theory.
This approach requires a uniform and homogeneous distribution of planetesimals both at the start and during 
a simulation. The issue is that in reality, the perturbation of the runaway growth bodies disturbs the 
distribution of planetesimals in their surrounding and breaks its uniformity \citep{Ida93,Tanaka94}\footnote{
The validity of the coagulation equation was later examined by \citet{ManHoi00}, as well.}.
In all above studies, either these non-uniformities were ignored, or the distribution of planetesimals was  
held uniform, artificially.

It is fundamentally important to note that the most realistic approach to studying planetesimal growth is simulating
their collisions by direct integration of their orbits. 
\citet{Aarseth93} were first to use this approach in the context of planetesimal growth\footnote{Prior to
this work, \citet{Cox80}, \citet{Lecar86} and \citet{Beauge90} used numerical interactions to simulate giant impacts
among planetary embryos.}.   
In a study on examining the predictions of their earlier work where
\citet{Palmer93} had presented an analytical treatment for the dynamical relaxation of a disk of planetesimals, these
authors integrated the orbits of 100 equal-mass and equal-size small bodies uniformly distributed around 1 AU.
We recall that triggering runaway growth requires non-uniformities in the size distribution of planetesimals, 
meaning that one or a few of these bodies must have larger sizes than those in their surrounding. 
\citet{Aarseth93} considered 
a disk of equal-mass and equal-size planetesimals (which they referred to as the least favorite environment for 
the runaway growth) to examine whether and how runaway growth would appear and proceed.
They showed that when allowing collisions with low relative velocities to result in coalescence
(larger relative velocities were considered to result in bouncing), some planetesimals do in fact show signs of
the runaway growth and increase their sizes to several times their original radii. However, because of the 
low resolution 
and short time of integrations (which were due to the limitations in their computational capabilities), 
the simulations by these authors did not show the sharp increase that is expected to be observed in the sizes of the 
runaway growth bodies. For that reason, their results stayed rather unnoticed. 

The first (relatively) high resolution simulations of runaway growth were carried out by \citet{Kokubo96}
and were later expanded by \citet[][see Kokubo and Ida 2012 for a review]{Kokubo98,Kokubo00,Kokubo02}. 
These authors simulated the growth of 3000 equal-mass planetesimals randomly distributed between 0.98 AU 
and 1.02 AU by integrating their orbits and allowing them to collide with one another. The initial eccentricities 
and inclinations of planetesimals were chosen following the distributions used in \citet{Ida92a}.
To avoid computational complexities due to the increase in the number of the bodies resulted from
the addition of collisional fragments, these authors considered growth by the way of perfect-merging.

As mentioned above, the onset of runaway growth requires one or a few planetesimals to be substantially larger 
than others. Because in the system considered by \citet{Kokubo96,Kokubo98,Kokubo02} all planetesimals 
had equal sizes, it would take a long time for their population to naturally produce a few large bodies on its own.
To speed up the integrations, these authors artificially increased the sizes of planetesimals by 5-10 folds 
at the beginning of simulations. \citet{Kokubo96} argued that, based on their approximated analytical method,
this artificial increase in the planetesimals sizes affects only the accretion time scale and plays no role on 
whether the runaway growth occurs or not.
The results by these authors showed that as planetesimals collide and grow, the locally large bodies undergo 
runaway growth in their surrounding, increasing their sizes by two orders of magnitude.
The perturbation from these bodies disturbs the orbits of planetesimals in their vicinity causing 
non-uniformities to develop in planetesimals distribution. \citet{Kokubo96} showed
that the runaway growth appears only when simulations are 3-dimensional, and results in the formation of 
planetary embryos within the first 20,000 years \citep[similar timescale has also been reported by][]{Kokubo98}. 
When co-planar systems are considered, growth through perfect-merging does not result in the 
runaway growth. Instead, similar to the results of \citet{Safronov69} and \citet{Nakagawa83}, it proceeds in an 
orderly manner.

The development of computational models capable of resolving collisions, combined with the advances 
in computer technology prompted researchers to improve the direct simulations of runaway growth by 
i) increasing the resolution of $N$-body integrations, ii) expanding the spatial range of simulations, 
and iii) including a certain number of collisional fragments. For instance, \citet{Richardson00} simulated 
the collisional growth of $10^6$ {\textit {uninflated}} planetesimals via perfect-merging in the region 
of 0.8-3.8 AU considering the perturbation of the present-day outer planets. These authors showed that, although
their integrations were so slow that it reached only 1000 years after 200 wall-clock hours, at which point they stopped
the integrations, some of the planetesimals still reached the state of runaway growth. However, due to
the short time of integration, the largest planetesimals did not reach the protoplanetary sizes and grew to only nine times 
their initial radii. To demonstrate the reliability
of their results, \citet{Richardson00} reproduced the results of \citet{Kokubo98} by integrating the orbits of 4000 
equal-mass planetesimals within the annulus of 0.96 AU to 1.04 AU. These authors inflated the initial planetesimals 
by 6 fold and showed that, in agreement with the results of \citet{Kokubo98}, their integrations were able to produce 
planetary embryos within the first 20,000 years. \citet{Kokubo02} also presented high resolution 
integration of planetesimal growth where they integrated the orbits of $(5-10) \times 10^4$ planetesimals distributed 
from 0.5 AU to 1.5 AU. These authors too increased the initial sizes of planetesimals by 6-10 folds and showed that when
the growth is modeled via perfect-merging, the first set of protoplanetary bodies with masses of 100 times larger than
the initial planetesimals appears in 50,000 years\footnote{We would like to
note that the runaway growth and planetesimal dynamics have also been studied by \citet{Rafikov03a,Rafikov03b} 
where the author presents an analytical treatment of the growth of an embryo in a disk of planetesimals, and
studies the dynamical evolution of the latter bodies. While, qualitatively, some of the results of these studies agree with the
results of previous numerical simulations, we do not discuss them here because some of their underlying 
assumptions ignore situations that appear in real systems.}. 

High resolution simulations of planetesimal growth were also carried out by \citet{Morishima08} and \citet{Barnes09}.
To examine the effect of remnant planetesimals on the orbital eccentricities and growth of planetary embryos,
\citet{Morishima08} integrated the orbits of up to 5000 equal-mass planetesimals in two annuli centered at 1 AU, one 
with a width of 0.3 AU and one with a width of 0.5 AU. These authors inflated the initial sizes of planetesimals by 
a factor of 4.3 and showed that when growth is modeled through perfect-merging, their systems reach the state of 
runaway growth at $\sim {10^4}$ years. 
\citet{Barnes09}  considered $10^5$ km-sized planetesimals at 0.4 AU, and integrated their orbits for 89 to 135 years
without inflating their initial sizes. These authors
demonstrated that, while some of the bodies did in fact grow by two orders of magnitude, runaway growth 
did not start as these bodies did not detach themselves from the rest of the population. We believe this 
lack of detachment (and with the same token, the non-appearance of runaway growth) is due to the short time 
of integrations by these authors.

The first $N$-body integrations of planetesimal growth that included collisional fragments were carried out by 
\citet{Leinhardt05} and were later followed by \citet{Leinhardt09}\footnote{We would like to note that the 
first time that fragmentation was implemented in an $N$-body integrator was in the work of \citet{Beauge90}. 
These authors simulated the last stage of terrestrial planet formation by integrating the orbits of
200 equal-mass fully formed embryos of 1.15 Moon-masses, and accounted for bouncing, cratering and fragmentation.}. 
These authors simulated the growth of $10^4$ 
planetesimals (initially inflated in size by 6 folds) in a 1 AU-wide area around the orbit of Earth and included 
collisional fragments derived from the collision  catalogs of \citet{Leinhardt00} and \cite{Leinhardt02}. 
Following those works, \citet{Bonsor15}, \citet{Leinhardt15} and \citet{Carter15} carried out similar simulations 
where, increasing the initial sizes
of planetesimals by 6 folds, these authors integrated the orbits of ${10^4} - {10^5}$ bodies in the region between 
0.5 AU and 1.5 AU. The growth in these studies were simulated using the accretion equation of \citet{Leinhardt05}, 
and collisional fragments were included using the analytic predictions of \citet{Leinhardt12} and \citet{Stewart12}. 
The results of all these studies broadly matched those of previous simulations demonstrating that the 
inclusion of fragments does in fact affect the evolution of planetesimals and runaway growth, although the effect may be minor.

Most recently, \citet{Clement20} studied the collision and evolution of planetesimals\footnote{It is important to mention  
that \citet{Carter20} also studied planetesimal growth. These authors integrated $10^5$ planetesimals 
allowing them to collide and grow following the models of \citet{Leinhardt12} and \citet{Leinhardt15} while subject to the perturbation 
of growing and migrating Jupiter and Saturn. Due to the short time of integrations (not longer than $10^5$ years), these authors did not 
inflate their initial planetesimals. We do not discuss the work of these authors any further as their results do not contribute 
to the topic of this paper and our study of runaway growth.}.
These authors considered five narrow annuli between 0.5 AU and 3 AU, and integrated the orbits of 5000 uninflated planetesimals 
in each annulus. Using an $N$-body integrator within a GPU 
environment, these authors showed that, when the effect of gas drag is mimicked by including the analytic gas-disk model of 
\citet{Morishima10}, runaway growth appears within the first 80,000 years. As we will demonstrate later, 
given the uninflated sizes of the planetesimals, this time of the appearance of runaway growth is too short, even
with the effect of gas drag included (gas drag will enhance the rate of growth).

Using a clever approach, \citet{Clement20} expanded their annuli integrations to the entire region of 0.48 to 1.65 AU 
(as one large and continuous area) and showed that, when the effects of giant planets are included, while as expected, 
this perturbation mainly affects the outer region of the asteroid belt, it is still large enough 
to have a moderate effect on the planetesimals orbits in the inner regions as well 
\citep[see][for more details on the effects of giant planets on planetesimals and planetary embryos in the terrestrial planet 
region]{Haghighipour12}. We will return to this topic and the work of these authors in more details in section 4.

At the time of this writing, the highest resolution simulations of planetesimal growth were those of \citet{Wallace19}. 
These authors integrated a set of low-resolution (4,000) and a set of high-resolution ($10^6$) planetesimals in a narrow
annulus between 0.94 AU and 1.04 AU for 20,000 years. In order to demonstrate the reliability of their integrations
by comparing their results with previous studies, these authors set the initial eccentricities and inclinations 
of their bodies similar to those in \citet{Ida92a}, and increased the initial planetesimal sizes by 6 folds. The 
20,000 integration time was chosen as the shortest time that would allow the authors to compare their results with those 
of \citet{Kokubo98} while their extremely slow high-resolution simulations would finish in a reasonable time.
Results showed that, in agreement with the work of \citet{Kokubo98}, simulations did in fact reach the state of runaway growth 
within $\sim 20,000$ years. They also demonstrated that the mean-motion resonances between growing planetary embryos 
(i.e., Moon- to Mars-size bodies) and planetesimals play an important role in the collisional growth of these bodies, resulting 
in the appearance of a bump in their final mass distribution \citep[see the bottom-left panel of figure 3 in][and the 
bottom-left panel in figure 4 of this study]{Wallace19}.

Before we continue, we would like to note that the studies of planetesimal accretion and
runaway growth have not been limited to the direct integration approach. Although $N$-body integrations present 
the most precise treatment of the growth and dynamical evolution of planetesimals, because realistic scenarios require 
integrating billions of bodies, which have been and still are beyond the capability of computational resources, 
many scientists studied planetesimal growth using statistical methods. For instance, 
\citet{Kenyon98,Kenyon99a,Kenyon99b}, \citet{Kenyon01,Kenyon02a,Kenyon02b}, \citet{Ohtsuki02} and 
\citet{Kenyon04a,Kenyon04b} 
used the particle-in-a-box approach of \citet{Safronov69} and studied planetesimal growth at the inner and outer
parts of the solar system, as well as in extrasolar planets. To overcome the breakdown of this approach due to the 
appearance of non-uniformities, \citet{Weidenschilling97}, \citet{Kenyon06} and 
\citet{Bromley06,Bromley11} developed hybrid  methods combining $N$-body integrations with an improved version of 
the coagulation scheme. Collectively, results of these studies showed that, contrary to direct $N$-body integrations,
runaway growth appears within 10-100 Myr, with this time increasing with the distance from the central star.

\citet{Ormel10a} and \citet{Ormel13} also studied runaway growth in the context of giant planet 
formation. \citet{Ormel10a} developed a statistical approach that would use a full Monte Carlo coagulation-fragmentation 
scheme, and showed that their approach preserves the individual nature of particles while it treats a large number of 
them statistically. In a later study, \citet{Ormel10b} used the results of this statistical approach and refined the 
conditions for transition from runaway growth to oligarchic growth. These authors showed that a state consistent with 
the runaway growth appears in the mass distribution of planetesimals at about $10^5$ years. 
\citet{Ormel13} developed a semi-analytical model
through which they studied the effect of dead-zones in turbulent disks on the onset of runaway growth. 
These authors found that
in the context of giant planet formation, the minimum size of a planetesimal to begin the runaway growth is 100 km
or larger, and that the growth of such bodies will overrun the life time of the nebular gas.

On the analytical fronts, \citet{Kobayashi10} studied planetesimal growth and developed a formula for the radius of 
runaway growth bodies. These authors defined their analytical radius in such a way that when used in studying planetesimal 
growth, the results would reproduce those of previous numerical studies, in particular those of \cite{Inaba01}.
Results of the numerical integrations by these authors demonstrated that, when considering growth via perfect-merging,
their defined version of runaway growth appears in $(1-4){10^5}$ years.
Using the statistical code of \citet{Inaba99,Inaba01}, \citet{Kobayashi10} also showed that their analytical 
results match those obtained from statistical analysis. The analytical and numerical approaches
of \citet{Kobayashi10} have been used in the study of debris disks \citep{Kobayashi14}, dynamical 
evolution of planetesimals
in gaseous disks \citep{Kobayashi15}, runaway growth in turbulent disks \citep{Kobayashi16}, and the growth of gas-giant
planets \citep{Kobayashi18}. Recently \citet{Walsh19} studied the growth of planetesimals to terrestrial planets
where they used tracer particles to simulate the growth of planetesimals and planetary embryos. 

Although the above-mentioned indirect approaches, as well as those prior to the work of \citet{Aarseth93} have greatly
contributed to our understanding of planetesimal growth, they are not discussed in this study. As mentioned earlier,
direct integration is the most realistic approach to planetesimal growth, and for that reason, it allows for studying
the underlying physical processes that actually affect the growth and evolution of planetesimals. We, therefore,
maintain our focus on direct integrations, using real-size planetesimals, and continue by examining the implications 
of the results for developing a realistic and self-consistent approach to modeling planet formation.

As mentioned before, the reason for inflating planetesimals is to shorten the integration time.
Although it has been known that this approach is not realistic, it has been argued 
that because, in general, enlarging planetesimals radii
does not disturb the effectiveness of gravitational focusing, it will not
affect the occurrence of runaway growth. In other words, as long as the study is qualitative, or for 
understanding the underlying physics of planetesimal growth, or for the purpose of proving
a concept, the error associated with inflating planetesimals may not be relevant. 
However, when the study is quantitative, and when the purpose is to develop formation models that are 
capable of making realistic predictions, the approach needs to be realistic. In such cases, unrealistic 
initial conditions will not allow simulations to portray the natural evolution of the system.
For instance, as shown in figure 1 of \citet{Bonsor15}, the time of growth increases by almost two orders of magnitude 
from $5 \times {10^4}$ years to 1.8 Myr for the appearance of the runaway growth, and from $2 \times {10^5}$ years to 7.2 Myr
for the formation of planetary embryos when the initial set-up is changed from using 6-fold inflated planetesimals to
no artificially expanded bodies. As shown by \citet{Levison03}, \citet{Haghighipour12}, and
\citet{Haghighipour16}, this time falls within the timescale that the gravitational perturbation of (growing)
Jupiter disturbs the protoplanetary disk, especially within the time that the secular resonance of 
Jupiter begin to appear and disturb the region around 1 AU. 

As the first step toward developing a realistic planet formation model,
it is fundamentally important to revisit runaway growth and examine its consequences using
initial conditions that allow for the natural evolution of the system. The rest of this article
has been devoted to this purpose. As mentioned earlier, our approach is to revisit runaway growth by redoing
some of the simulations and analyzing the results in connection with those in the literature. 
When carrying out simulations, we maintain focus on principle concepts and take a conservative approach: We consider a system 
consisting of equal-size and equal-mass planetesimals.
It is understood that such distribution of planetesimals is not entirely natural (a natural distribution
will include planetesimals of different sizes and masses). However, quoting \citet{Aarseth93}, as ``the least favorable environment
for runaway growth'', such a system will give us a less biased view into the process of the formation
of planetary embryos.
Also, to avoid complications due to the proper handling of collisions, we consider
growth through perfect-merging. Although unrealistic, this scenario presents the most efficient mode of 
growth meaning that, any conditions and requirements that appear when using perfect-merging will only be
enhanced when the collisions between planetesimals are resolved realistically.
We refer the reader to \citet{Haghighipour22} for a detailed analysis of the errors due to perfect-merging, and the remedy for them.
Finally, we do not enlarge any planetesimal at the beginning of the integrations and
to maintain focus on the underlying physics of runaway growth and its timescale, we do not include additional physical 
processes such as the perturbation of the growing giant planet(s) (these effects are included in a subsequent study), 
the damping effect of the nebular gas drag, the increase in the number of bodies due to fragmentation, and the dynamical 
friction due to the collisional debris (the dynamical friction due to the bi-modal mass of the disk is implicitly included).

\section{Revisiting planetesimal growth}
\label{section:initial_setup}

As mentioned earlier, to identify the requirements for a realistic model, the foundations of planet formation have to be
revisited. Because in the context of planetesimal growth, many of the fundamental studies are more than 25 years old, and because 
those studies were carried out in contexts other than identifying requirements for realistic models, we revisited this topic 
by performing those fundamental simulations, but without inflating planetesimals. In the rest of this article, we will present 
the results of these simulations in the context of all previous investigations and discuss their implications both in connection 
to one another and in connection with previous studies.

As in many of the previous $N$-body simulations of planetesimal growth, we followed \citet{Kokubo96} and 
randomly distributed approximately 3000 equal-mass planetesimals around 1 AU. Several studies, such as those cited here,
have demonstrated that a population of 3000 - 4000 planetesimals would be sufficient to reveal and study the underlying
physics of planetesimal growth without overloading the integrations. We set the 
mass of each planetesimal to $m={10^{23}}$ g and its density to 2 g cm$^{-3}$. The total mass of the disk was 0.05 Earth-masses.

In their simulations, \citet{Kokubo96} distributed
planetesimals between 0.98 AU and 1.02 AU. These authors argued that, based on the results of \citet{Wetherill89}, 
this width is large enough for the planetesimals to stay within its boundaries for the duration of the 
integrations. To examine the effect of the spatial distribution of planetesimals on their growth, we 
followed similar approach and considered two sets of simulations, one with an annulus from 
0.96 AU to 1.04 AU (hereafter, set A) and one with an annulus extending from 0.98 AU to 1.02 AU 
(hereafter, set B). In each set, we ran five different simulations.

As mentioned earlier, runaway growth is triggered when simulations are carried out in 3D. We, therefore,
considered a Gaussian distribution\footnote{We would
like to note that strictly speaking, the Gaussian distribution applies to each component of the
random velocity. However, the norm of the random velocity, in our simulations, the eccentricity, 
follows a Rayleigh distribution.} for the eccentricities and inclinations
of planetesimals with a zero mean and a dispersion of $<e^2>^{1/2} = 2 <i^2>^{1/2} = 2h$. 
Here $h={R_h}/a$ is the reduced Hill's radius \citep{Ida92a} and

\begin{equation}
 {R_H} = \left( \frac{m}{3M_\odot}\right)^{1/3} a
\end{equation}

\noindent
is the Hill radius of a planetesimal. In equation (2), $m$ and $a$ are the planetesimal's mass and semimajor axis,
respectively, and $M_\odot$ is the mass of the Sun. The arguments of the pericenter and 
mean anomalies of planetesimals were randomly chosen using a uniform distribution, and the longitudes of 
their ascending node were set to zero.

We integrated each system for $5 \times {10^5}$ years using the hybrid routine in the $N$-body integration package
MERCURY \citep{Chambers99}. The time-steps of integrations were set to 6 days\footnote{When using symplectic 
integrators similar to the hybrid routine in the MERCURY package, it is recommended to set the integration time-step
to smaller than $1/20^{\rm th}$ of the shortest orbital period in the system. Our time-step is equivalent to 
$1/60^{\rm th}$ of the shortest orbital period at the start of the integrations}. As stated above, 
integrations did not include artificial inflation of planetesimals. We considered two objects with masses $m_1$ and $m_2$
on a possible collision course when they were closer than three times their mutual Hill's radius $({R_{H_{12}}})$, 

\begin{equation}
{R_{H_{12}}} = \left(\frac{{m_1}+{m_2}}{3M_\odot}\right)^{1/3} \left(\frac{{a_1} + {a_2}}2\right)\,,
\end{equation}

\noindent
and, in cases when the objects collided, we simulated collisions as perfect-merging. In all integrations, the total energy 
and angular momentum were conserved with a relative error of $10^{-10}-10^{-11}$.

\section{Results and comparison with previous studies}

Table 2 shows the masses and growth times of the three largest embryos at the end of each integration.
As shown by the column ``Final Mass''\footnote{We would like to note that by ``Final Mass,'' we refer to the 
mass of the object at the end of the integration.}, the masses of these bodies vary between 20 and 142 times the 
mass of the initial planetesimals. That means, not only did runaway growth occur in all our simulations, 
in most systems, the runaway bodies continued their growth even up to the boundary of the oligarchic 
regime\footnote{Oligarchic regime is reached when the runaway growth bodies become so large that the growth 
becomes exclusive to a few protoplanets (the oligarchs). At this stage, these bodies grow slowly to their 
final sizes with their mass-ratios approaching unity. During this time, the dynamics of the planetesimals
is no longer the result of their mutual interactions, but is governed by their interactions with the growing 
embryos (the oligarchs).}.

Figures 1 and 2 show the snapshots of the evolution of a sample of our systems. In figure 1, we show 
systems A1 and B4 as examples of those in which the mass of the largest embryo is more than 100 times the initial
planetesimals, and figure 2 shows systems A2 and B5 as samples of the rest of the systems. 
Each object is shown by a circle with its radius proportional to the object's mass. Blue circles represent the 
bodies with masses at least 20 times their initial masses. 
The black circles in the bottom panels show the three largest bodies at the end of the 
integration. The black circles in prior panels show the same three bodies as they grew and evolved during 
the simulation.

\subsection{Runaway growth}
As shown by figures 1 and 2, runaway growth has occurred in these (and by the same token, in all our) 
simulations. We show the latter using two approaches. First, we use the definition of runaway growth as
presented by \citet{Kokubo96}. We will then explain the inconsistencies in using this definition and demonstrate 
the occurrence of runaway growth by using a more accurate criterion based on the analysis of the distribution of 
masses after a runaway growth system has reached relaxation (i.e., when its mass distribution follows a power-law;
see equation 4 and figures 4 and 5).

\citet{Kokubo96} defined runaway growth as when 1) the largest object (i.e., the runaway growth body) grows locally 
(i.e., within its accretion feeding zone)
faster than the second largest body, and 2) that the ratio of the mass 
of the runaway growth body to the mean mass of the rest of the system increases with time. We show in figure 3 
the growth of all large bodies of figures 1 and 2, as well as the time-variation of 
the mean mass of each system when the largest body is {\textit{not}} included. 
The top panels correspond to 
systems A1 (left) and B4 (right), and the bottom panels are for systems A2 (left) and B5 (right), respectively.
In each panel, red, green and blue curves represent the growth of the three largest bodies (the black circles
in figures 1 and 2) and the gray curves correspond to the growth of all blue circles in those figures (i.e.,
all other objects with a mass larger than 20 times the initial planetesimals). 
As shown here, all large bodies grow much faster than the mean mass of their respective systems, 
fulfilling the above criteria for runaway growth. We have observed similar trend in all our simulations.

Although the above definition of runaway growth can explain the formation of our runaway bodies, 
it suffers from a fundamental inconsistency:
The term ``locally'' in this definition refers to the local feeding zone of an object and requires this zone to be
static. However, as an object grows, its position changes with time and so does its local feeding zone. 
This can be seen in figures 1 and 2 where
the semimajor axes, and therefore, the spatial locations of the growing bodies (e.g., the black circles) change
during the integration.
This change of the orbit and mass 
causes the location and size of the feeding zones of the bodies to change as well, making it practically
impossible to determine the feeding zone and the second largest body to which the above
definition of runaway growth applies. As noted by \citet{Kokubo00}, and as explained below, 
a more accurate way of demonstrating whether runaway growth has occurred would be 
to study the mass-distribution of the system when it has reached relaxation.

\subsection{Mass Distribution}
To study the mass-distribution, we note that, as shown by 
figure 3, at the end of an integration, the mean mass of the rest of the system only doubles.
In other words, most of the remaining mass is still in the form of the initial planetesimals 
(as we explain below, this is another characteristic of a system in which runaway growth has occurred). 
We demonstrate the latter in figures 4 and 5 where we show the time evolution of the distribution of mass 
(that is, the number of objects for different values of their masses) in systems of figures 1 and 2. Each point 
in these figures corresponds to the number of planetesimals with the same mass. 
Note that, because growth has been modeled through perfect-merging and because, initially,
all planetesimals had equal masses, the growth of an object will appear as the multiples of its initial mass.

As shown by figures 4 and 5, there is a strong correlation between the number of bodies and their masses up 
to 10 times the initial mass of individual planetesimals (i.e., up to $10^{24}$ g.) 
The mass distribution in this interval seems to follow 
a power law of the form

\begin{equation}
N(m) = N_1 \,  m^{-\alpha},
\label{eq:nvsm}
\end{equation}

\noindent
where $N_1$ is the number of the planetesimals remaining from the initial population. 
As shown by \citet[][and references therein]{Kokubo96,Kokubo98,Kokubo00}, in systems where runaway growth occurs, 
$\alpha > 2$. We show the latter in Table 3 where we have listed the values of $N_1$ and $\alpha$ 
for all our systems at the end of their integrations. As shown here, $\alpha$ stays larger than 2 
for the entire time during integrations confirming that runaway growth occurred in all our simulations.

Because equation (4) shows the distribution of small masses, if, while the objects grow (for instance, through 
perfect-merging), there is no re-supply of small bodies, the value of $\alpha$ will decrease with time.
\citet{Kokubo96} showed that in their simulations (where the planetesimals had been inflated), 
$\alpha$ decreased from 2.6 to 2.4 while maintaining an 
averaged value of $\sim 2.5 \pm 0.4$. \citet{Kokubo00} reported that at the end of their simulations (where 
planetesimals had not been inflated), the final value of $\alpha$ varied between 2.2 and 1.9. In our simulations, 
$\alpha$ exhibited similar behavior. For example, in the systems of figure 4, $\alpha=2.57$ in the top-left panel 
and 2.6 in the top-right panel, and in the systems of figure 5, these values are 2.8 and 2.6 for the top left and 
right panels, respectively. During the course of the integrations, these values dropped to those in Table 3,
which are in stark agreement with the values reported by \citet{Kokubo00}.

It is important to note that the mass distribution portrayed by figures 4 and 5, and formulated by
equation (4), while the direct consequence of the growth via perfect-merging, is independent of radius inflation. 
Because when a planetesimal is inflated, it is only its radius that is enhanced (and its mass is not changed), 
the planetesimal's bulk density will be reduced by the same factor as its radius enhancement. Since the mass of
the perfect-merging body is equal to the sum of the masses of the two colliding objects, the decrease in the bulk 
density of the impactors cancels out the effect of their size inflation, leaving their mass distribution intact.

However, as mentioned earlier, perfect-merging is unrealistic. In reality, collisions result in break-up,
fragmentation, and shattering. Each of these events affects the number, as well as the mass and size distributions of 
the bodies, differently. The latter implies that, although the appearance of the runaway growth is independent of the 
modes of collision and growth (as its occurrence is the result of the appearance of non-uniformities in the planetesimals 
population), the time of its appearance, its duration, and the mass and size distributions of the resulted bodies
will be different from those obtained from perfect-merging scenario. Unfortunately, at present, it is not possible
to determine mass-distribution realistically as computational facilities do not have the necessary capabilities to 
resolve collisions and growth among tens of millions of objects, accurately.

\subsection{Remaining small bodies}

An inspection of figures 4 and 5 indicates that during the evolution of a system, a large number of small bodies 
(in the form of both initial planetesimals and objects a few times more massive) stay in the system. 
Figure 6 shows this by demonstrating the decrease in the number of objects in each system. We also show the number of 
the remaining bodies at each time of the integration on the upper left corner of the panels in figures 
1 and 2. As shown here, the number of small
bodies at the earlier stage of integrations, for instance at $t=125,000$, where the value of $\alpha$ is between
2.8 and 2.5,  is even larger indicating that when runaway growth is dominant, most of the 
mass is still in the form of small objects (see the graph of the mean mass in figure 3). 
 
These remaining small bodies play a vital role in the subsequent evolution of the system. The collective effect
of these objects increases the
efficacy of collisional growth by damping the orbital eccentricities and inclinations of growing bodies 
through dynamical friction. We have demonstrate the latter in figures 7 and 8 where we show the RMS values 
of the eccentricities (filled circles) and inclinations (open circles) of the bodies of the systems of figure 1 and 2. 
As shown by the top panels, at the early stage of the evolution, when objects 
have just started to grow, their eccentricities and inclinations are strongly damped. This trend is maintained 
throughout the integration by small bodies ranging from 1 to 10 times the mass of the initial planetesimals 
(shown by $m_{\rm min}$). As larger bodies appear, their perturbation causes the eccentricities and
inclinations of all objects to increase. The latter can be seen for the value of the mass larger than 
$10 \, {m_{\rm min}}$ and is more pronounced in figure 7 which corresponds to systems A1 and B4. In these 
systems, embryos are larger than the systems of figure 8 (A2 and B5, see Table 2) with the largest embryos being 
1.5 to 3 times more massive, causing the eccentricities and inclinations of smaller bodies to raise
to slightly higher values.

Similar results have  been reported by other researchers, though with some differences. 
For instance, at the end of the simulations by \citet{Kokubo96}, 65\% of the initial 
planetesimals were accreted by the runaway growth bodies. \citet{Kokubo00} report this number at 56\%. These
authors carried out simulations for 200,000 years. In our 
simulations, the number of accreted bodies ranged between 48\% and 57\%.  We believe that the larger number
of accreted bodies in the work of \citet{Kokubo96} is due to the inflating of planetesimals at the beginning of
simulation, and is, therefore, unrealistic. 
When realistic sizes are used, as in our simulations and those of \citet{Kokubo00}, the collision 
cross-section is not artificially enhanced and as a result, the number of accreted bodies drops. The larger value of
this number in the simulations of \citet{Kokubo00} can be attributed to the effect of nebular gas drag.
As explained in the Appendix, the drag force of the nebula operates as an additional factor in damping the 
eccentricity which subsequently results in lowering the relative velocities of colliding bodies.

We would like to emphasize that because accretion is a function of time, it would, generally, be more meaningful to 
compare the rates of accretion in different simulations at the same times. However, in this study, that is unnecessary as 
our goal in the above paragraph is to merely demonstrate that the high rate of accretion in simulations with inflated 
planetesimals is in fact artificial and, therefore, such high accretion rates and any results obtained from them
must not be considered as the natural evolution of the system.

An interesting result depicted by figures 7 and 8 is the manner by which the small material at the end
of the simulations reach their final eccentricities and inclinations. As shown here, in systems of figure 7,
the bottom panels show an almost constant distribution for the objects in the
1 to 10 mass-range. 
However, in the systems of figure 8,
where the final embryos are smaller than those of figure 7, objects in the same mass range continue their 
linear trend toward lower values of eccentricity and inclination. Note that the trend in the system on the right in 
figure 8 (system A5) is weaker as in this system embryos are larger than those in system A2 shown by the left column. 
It is important to note that in these simulations, the dynamical evolution of the large bodies cannot be 
studied statistically (meaning, their evolution cannot be fitted or used to identify a trend) as the number of 
these objects is too small for the statistical analysis of their dynamics to produce meaningful results.

\section{Implications of the results}

In the last section, we presented the fundamental characteristics of a system of planetesimals within a narrow
annulus in a protoplanetary disk. We showed what it means when the runaway growth occurs, and how the system evolves and 
responds to it. While most of the results can be found in previously published articles, we presented them here collectively, 
so that the reader could see the connection between them at one place. We also demonstrated how these results were obtained 
by carrying out relevant simulations. In this section, we follow the same approach and present the implications of the
results. We discuss how specific initial conditions manifest themselves in the final outcome, and what the characteristics of 
the system suggest for carrying out realistic simulations (i.e., simulation with no ad hoc, unrealistic, and/or 
simplifying assumptions).

In general, the results presented in the previous section 
have four implications for the late stage of terrestrial planet formation. 
They show that 1) when planetesimals are not inflated, planetary embryos do not reach the high masses that simulations 
with inflated planetesimals suggest. That means, 2) in actual systems, the evolution of the protoplanetary disk and 
the process of planetesimal growth are subject to the perturbation of embryos smaller than those obtained from these 
simulations. They also imply that
3) the growth of embryos takes much longer to the extent that while they go through the 
runaway and oligarchic modes, their dynamical evolution and, therefore, their isolation masses and orbital 
architecture may be affected by the growing giant planets. Finally, results show that while simulating planetesimal
growth using isolated distributions (e.g., a planetesimals annulus) provides an excellent
approach to understanding the mechanics of the process and its underlying physics, 4) in order to develop
realistic initial conditions that can be used in simulating the last stage of terrestrial planet formation, 
simulations of planetesimal growth have to be carried out for the entire of a planetesimal disk at the same time. 

In the following we explain each of these implications in more detail.

\subsection{The effect of inflating planetesimals}

A comparison between the results of the last section and those in which planetesimals were
initially inflated immediately demonstrates that our runaway growth bodies are, in general, much smaller than in 
those simulations. For instance, in the simulations of \citet{Kokubo96} and \citet{Richardson00}, after only 
20,000 years of integrations, the runaway growth bodies reached the masses of 300-400 time the mass of the initial 
planetesimals $(m_{\rm min})$. In the works of \citet{Leinhardt05} and \citet{Wallace19}, these values rose up to 
1000-1500 $m_{\rm min}$ during the same time. In our simulations, however, after 500,000 years of integrations, 
the masses of our runaway growth bodies reached only 20-142 $m_{\rm min}$ with our largest embryos having a mass 
between 107 $m_{\rm min}$ and 142 $m_{\rm min}$.

Similar results have also been reported by \citet{Kokubo00} and \citet{Clement20}. These authors did not initially
inflate the planetesimals, and also included the effect of gas drag. In the simulations of \citet{Kokubo00}, it took 200,000
years for the largest planetesimals to reach a mass of 200 $m_{\rm min}$, and in the work of
\citet{Clement20}, the largest body in their annulus centered at 1 AU reached 550 time of its initial mass after 100,000
years. It is important to note that the shorter time of growth in the work of \citet{Kokubo00} is due to the effect of gas
drag, and that in the simulations of \citet{Clement20}, the time of growth is even shorter is due to the fact that
these authors used almost twice as many planetesimals in their 
1 AU centered annulus as those in ours and in the simulations of \citet{Kokubo00}.

These small sizes of runaway growth bodies in simulations with uninflated planetesimals, combined with the value 
of the exponent $\alpha$ in each of our systems (see Table 3), and the fact that in such simulations, integrations 
had to be carried out for 5 to 25 times longer than those with inflated planetesimals, strongly implies that in 
simulations where planetesimals are not inflated, the dynamical evolution of the 
system would follow a path in which it would naturally take much longer for planetary embryos to reach their 
isolation masses and for the growth 
to transition from runaway $(\alpha \geq 2)$ to oligarchic mode $(\alpha < 2)$. 
For instance, an examination of Table 3 indicates that while systems A1, B4 and B5 are at 
the verge of transitioning to oligarchic growth, other systems still require more time, some of them 
(e.g., systems A2, A3, A5) much longer than our integrations time of 500,000 years. 

The fact that in systems with inflated planetesimals, large bodies form in much shorter times indicates
that from the early stages, the dynamical evolution of these systems, as well as the growth and orbital
architecture of their bodies are heavily influenced by the perturbation of their unrealistically
large embryos. We note that the large sizes of runaway growth bodies in simulations with inflated planetesimals 
is an expected result that has roots in the fact that 
increasing the initial sizes of planetesimals enlarges their collision cross sections and increases the rate of 
their growth causing very large bodies to form at short times.
As a result, soon after the start of the simulations, these systems evolve along a path that is dictated to them by the 
perturbation of their unrealistically large bodies. This unreal evolution propagates to all stages 
of planetesimal growth and embryo formation making the spatial and mass distributions of these bodies 
unreal as well. The latter
will affect the final stage of terrestrial planet formation causing the system to produce planetary masses 
and orbital architecture that are inherently contaminated by the unnatural evolution of their original systems.

We would like to note that, in general, a better comparison between the results of two simulations would 
be when the ratio of the number of their remaining bodies to the number of their initial bodies are similar. 
For instance, the number of remaining bodies at the end of the integrations of \citet{Kokubo96} 
$(\sim 35\%)$ is about $6.7\%$ to $17\%$ smaller than the final 
bodies in our simulations implying that the results would have been better compared if our integrations had been
continued till the number of the remaining bodies in our simulations would reach $35\%$. However, for the purpose of this
article, such extension of integrations would be unnecessary. As shown by Table 2, given the 
rate of the planetesimal growth in our system, it is very unlikely that the masses of our runaway growth bodies
would have reached 300-400 initial planetesimals even if we had continued the integrations. Even if 
the effect of nebular gas drag had been included, given the similarity between our initial planetesimals population
and those of \citet{Kokubo00} who considered gas drag, our planetesimals would have at most reached the maximum mass of 
200 $m_{\rm min}$ as obtained by these authors. That, however, would have had no significant qualitative 
impact on the implications of the results. In other words, while extending integrations to longer times would have 
allowed for a quantitatively more accurate comparison with previous works, the implications of the results, as stated 
in this section and summarized in Section 5, stay intact.

\subsection{The effect of the spatial distribution of planetesimals}
As shown by Table 2, at the end of the integrations, with the exception of system A1, the masses of 
the three largest bodies in the systems of set B are much larger than those in set A.
This is an expected result that is the consequence of the initial distribution of planetesimals.
As mentioned in Section 3, in the simulations of set A, planetesimals were initially distributed over a range of 
0.08 AU whereas in set B, same number of planetesimals were scattered over 0.04 AU. The latter concentration of 
planetesimals in a smaller region naturally increased the rate of their collisions, and because collisions were 
taken to be perfectly inelastic, resulted in their growth to larger objects. This can also be seen in the work of
\citet{Kokubo00} where the initial distribution of planetesimals was over a smaller region (0.02 AU) and as
a result (and also because of the effect of gas drag), their largest body was even larger 
than those in set B (200 times the mass of the initial planetesimals)\footnote{We would like to caution that a full 
comparison with the results of \citet{Kokubo00}, and attributing the differences in results presented here to a single 
cause, namely the distribution of the planetesimals is not warranted as there are many differences in the setup used 
by these authors and ours. For instance, \citet{Kokubo00} considered gas drag, used a boundary condition in which the
loss of a planetesimal at one edge of the annulus due to the expansion of the disk was compensated by artificially introducing
a new planetesimals at the other edge, and used an enhanced surface density where their disk was 50\% more massive than
minimum-mass solar nebula.}. In contrast, when the same number of 
planetesimals were distributed in a larger region (e.g., set A), the surface density of their 
annulus was smaller, which, because initial planetesimals were equal-mass, resulted in their number density 
to be smaller as well. A smaller number density resulted in a smaller number of collisions, and therefore, 
less massive bodies 
\citep[see, e.g., the explanation about over-populated annuli as used in][in the second paragraph of section 4.1]{Clement20}.

\subsection{Expansion of initially localized distributions}

It is important to note that during the evolution of a system, any local distribution of planetesimals will
gradually expand. Figures 1 and 2 show that by the end of the integrations, the range of the 
planetesimals' semimajor axes has expanded by about 0.04 - 0.06 AU. Similar expansions can also be found in
the works of \citet{Kokubo96}, \citet{Richardson00}, \citet{Morishima08}, and \citet{Wallace19}. For instance,
at the end of the simulations in \citet{Kokubo96}, the semimajor axes of planetesimals expanded by 0.06 AU. 

In general, this expansion is caused by 1) diffusion due to the mutual interaction among planetesimals (as in our 
simulations), and 2) in cases where the drag effect of the nebular gas is included, by gas drag which 
causes planetesimals to migrate toward the central star (known as gas drag-induced migration). 
When the initial distribution of planetesimals is localized, this expansion reduces planetesimals surface density 
and after a while, impedes further formation and growth of large bodies. 

While diffusion and gas drag-induced migration are real physical processes that occur
naturally during the dynamical evolution of planetesimals, the delay and/or the disruption
of the growth is an artificial effect that appears because of the decrease in the surface density
which itself is a consequence of using localized distributions.
In a real system, where the initial extent of the planetesimals disk is over a much larger region, 
when a local distribution expands, its neighboring distributions also expand. As a result, while a local distribution
is losing bodies due to migration and diffusion, planetesimals from its neighboring distribution migrate/diffuse into it, 
compensating for the ones that were lost. The latter prevents the decrease in the surface density of the disk and 
allows the growth to continue. It is important to note that because the mechanics of the expansion is different for 
disks with different surface densities, and because that is what determines how neighboring regions spill into each other,
the efficiency of this mechanism is heavily governed by the disk's initial surface density profile 
\citep[see e.g.,][]{Izidoro15,Haghighipour16}

The fact that in a real disk, planetesimals from one local distribution enter their neighboring units prompted \citet{Kokubo00} 
and \citet{Clement20} to account for the reduction of the surface density by introducing a boundary condition in which after a 
planetesimal has left one boundary of a local distribution, that planetesimal is removed from the integration and a new one with 
the same mass is added at the opposite boundary. In the simulations of the last section, we chose not to adopt this approach because
while on the surface, this boundary condition remedies the loss of material and reduction of surface density, it suffers from the 
following shortcoming: The effect of an object artificially added to one boundary is not identical to the effect of a body that diffuses 
out of the system in the opposite side. Also, and more importantly, in the simulations presented here, our main intention has been 
to reveal the issues associated with using isolated distributions, and that reliable results cannot be obtained using this type of systems.
Finally, as we explain in section 4.3.1, unless integrations are expanded to the entire disk at the same time, any workaround,
especially using a collection of isolated distributions will still suffer from above issues and internal inconsistencies.

\subsubsection{Simulating the entire disk at the same time}

That the decrease in the disk surface density, and the subsequent impeding of the growth are artifacts of using a localized 
distribution strongly implies that simulations with localized mass distributions, or a collection of localized mass distributions 
(i.e., rings), do not produce reliable results, and in a realistic simulation, the entire disk of planetesimals 
needs to be integrated at the same time.

At the time of this writing, the only attempt in extending integrations to the entire of a disk was due to \citet{Clement20}.
As mentioned in the Introduction, these authors simulated planetesimal growth in five separate rings between  0.5 AU and 3 AU 
(see Table 1 for the locations of their planetesimal annuli). To extend their isolated simulations to a larger region  
(0.48 -- 1.65 AU), the authors made the assumption that any ring-like distribution of planetesimals between each 
two of their main annuli, when integrated, would, in general, exhibit the same dynamical behavior as those in their 
five main rings, and produce similar results. With that assumption, the authors divided the region of 0.48 -- 1.65 AU 
into small rings and interpolated the results of their simulations into these new areas. 

It is important to note that the simulations of planetesimal growth are extremely stochastic. That means, they 
are unpredictable and cannot be regulated. This lack of predictability implies that in principle, it is not possible 
to ensure that the dynamical evolution of planetesimals in different distributions, even if geometrically similar, 
would be identical, or carry certain predictable features. The interpolation approach used by \citet{Clement20}, 
although undoubtedly clever and most likely useful in demonstrating some proofs 
of concepts, is subject to this unpredictability meaning that, its results, may not be the true representation of 
the dynamical evolution of planetesimals, if used in lieu of the direct integration of the entire system. In other 
words, to obtain a realistic image of the dynamical evolution of planetesimals, the entire disk has to be integrated 
at the same time.

\subsubsection{A note on recent efforts on simulating planetesimal growth in ring-like distributions} 

Despite the above-mentioned effects of the expansion of isolated distributions on the growth and dynamics of planetesimal,   
in the past few years, a series of articles have promoted planetesimal growth in rings as a way of 
explaining the formation and orbital architecture of terrestrial planets in our solar system 
\citep{Ogihara18,Broz21,Izidoro22,Morbidelli22}, as well as the formation of super-Earths in extrasolar planets  \citep{Batygin23}. 
In these studies, converging radial migration of solids in the gaseous component of the disk has been presented as the way of accumulating 
these bodies in ring-like distributions. This radial migration has been attributed to the appearance of pressure enhanced regions due to 
the interaction of stellar wind with MRI active/inactive regions of a disk \citep{Ogihara18}, the mere assumption of a pressure enhancement 
near the CO snowline, water snowline, and silicate sublimation line \citep{Izidoro22,Morbidelli22,Batygin23}, and to the Lindblad, co-rotation, 
and heating torques applied to a protoplanet due to its interaction with the gas. These studies suggest that the accumulated bodies 
in such regions can grow through collision as well as accretion of pebbles, and form larger bodies including terrestrial planets and super-Earths. 
Recently \citet{Kambara25} have shown that it is also possible for runaway growth bodies to undergo oligarchic growth while in such isolated
distributions, and as their distributions expand. 

We would like to note that while the underlying physics of converging gas drag-induced migration in the vicinity of pressure enhanced regions 
is solid and fully supported by the physics of solid-gas interaction \citep[we refer the reader to][for the full theory of dynamical evolution 
of solid objects in the vicitiy of gas pressure/density enhancements]{Haghighipour03a,Haghighipour03b}, and although the appearance of Lindblad
and co-rotation torques can indeed affect the orbits of protoplanetary bodies, caution must be taken in using results of these simulations as 
these models include assumptions that have been tailored to facilitate the goals of their studies. For instance, in the simulations of 
\citet{Ogihara18}, the accumulation of planetesimals at the pressure enhanced regions and their subsequent growth occur only if the planetesimals 
are km-size and no larger than 10 km. The process fails when the objects are larger. In contrast, in the work of \citet{Broz21}, 
torques are more efficient in accumulating bodies when the objects are large, as the interaction of these bodies with the 
gaseous disk needs to be strong so that the resonance density waves and their resulting spiral arms can produce strong Lindblad and co-rotation
torques. Finally, in the works of \citet{Izidoro22,Morbidelli22} and \citet{Batygin23}, the pressure enhancements have merely been assumed, 
and their associated gas drag-induced force has been included in the equations of motion, analytically.

\subsection{The effect of the long time of the runaway growth}
As shown by figures 1, 2 and Table 2, for any given mass, the time of growth in our simulations is considerably long.
For instance, after almost $(4.5 - 5) \times 10^5$ years, the five largest embryos in all our
simulations grew to only 92 - 142 times the initial planetesimals $(m_{\rm min})$, and the three largest bodies 
in each system grew only slightly larger than $100 \, m_{\rm min}$.
This long formation time is a characteristic of realistic simulations and is the main reason that many researchers
opted for inflating planetesimals. \citet{Kokubo96} showed that when the gravitational focusing is
effective, an increase in the initial sizes of planetesimals by a factor $f$ reduces the time of the 
formation of the runaway growth bodies by a factor between $f$ and $f^2$. 
These authors inflated planetesimals by a factor $f=5$ and their runaway growth bodies  
reached the mass of 400 times the initial planetesimals (almost 4 times the masses of our final bodies) 
in $\sim 20,000$ years, a time that is $5^2=25$ times shorter than the growth time of the five largest embryos 
in our simulations.

Similar long formation times have also been reported by \citet{Kokubo00}. Recall that these authors did not 
inflate the initial planetesimals. The results of the simulations by these authors showed a runaway growth body 
with a mass 200 times the initial planetesimals forming within $2 \times 10^5$ years. Although this time is shorter than the 
formation time in our simulations, it is still 10 times longer than the time of the formation of runaway growth 
bodies in simulations of \citet{Kokubo96} and \citet{Richardson00}. Other simulations such as those by 
\citet{Kortenkamp01}, \citet{Carter15} and \citet{Wallace19} also hinted at long formation times. For instance, 
\citet{Carter15} and \citet{Wallace19} simulated planetesimal growth using inflated planetesimals and stated that 
in order to produce similar results using uninflated planetsimals, they would have had to continue integrations to
at least 750,000 years or more likely to a few million years.

Given that runaway growth is an inevitable consequence of the stochastic nature of planetesimals collisions, 
the timescales presented in our simulations, supported by their agreements with theoretical predictions and results 
of similar studies, strongly suggest that, in real disks, the time of the formation of runaway growth bodies
is most likely of the order of a few to several hundred thousand years and maybe even longer. These long timescales,
fall within the time of the growth of the cores of giant planets and the process of their 
gas-accretion. As shown by \citet{Haghighipour12}, in our solar system, the growth of giant planets strongly 
affects the dynamics of small bodies interior to the orbit of Jupiter. 
It is, therefore, imperative that the simulations of planetesimal growth
in the inner region of the Solar System protoplanetary disk to be carried out simultaneously and 
in concert with the simulations 
of the growth of giant planets in the outer regions. \citet{Carter15} have presented such calculations in the 
context of the Grand Tack model. However, their study suffers from two shortcomings. First, these authors used inflated 
planetesimals, and second, as shown by \cite{Morbidelli18}, Grand Tack model is inconsistent with 
the compositional properties of the moon and terrestrial planets, implying that the problem is still open 
and simulations needs to be carried out with giant planets forming in or in the vicinity of their current orbits.

\section{Summary and Conclusions}

In this article, we presented the requirements that simulations of planetesimal growth need to fulfill in order to 
portray a more realistic image of the early stages of planet formation, and to produce physically meaningful and reliable 
results. To place the analyses in the right context, and to maintain focus on the basics, we started by an
in-depth review of the history of the field where we discussed different approaches to modeling the growth of planetesimals, 
as well as their advantages and shortcomings. We focused our analysis on direct, $N$-body integrations as these integrations
present the most precise approach to planetesimal growth. We identified artificially increasing planetesimals' sizes at the start of
simulations, a practice that has been used by many researcher for the past three decades for the mere purpose of reducing the 
integration times (see Table 1), as the most unrealistic aspect of these integrations. We showed how inflating planetesimals 
affects the results, 
and how its unrealistic outcome extends to the subsequent phases of oligarchic growth and giant impacts.

With the goal of identifying the requirements for a realistic approach, we discussed the results of the simulations of planetesimal 
growth by carrying out such simulations in a localized environment without inflating their initial sizes. 
We identified the above requirements and demonstrated how they emerge from the natural evolution of the system 
by analyzing the implications of the results in connection with similar simulations in the literature.
To maintain focus on the underlying physics of runaway growth and the dynamical characteristics of the system, 
we considered collisions to be perfectly inelastic and did not include additional physical processes such as gas drag, 
fragmentation, and the effect of debris. It is understood that neglecting some of these processes, in particular, gas drag, 
may not allow the results to be fully realistic. However, comparisons with previous works indicate that the lack of these 
processes has only moderate quantitative effects on the final results, and do not change the nature of their implied constraints 
and requirements, qualitatively.

The following presents a summary of the analyses and their implied requirements for any realistic model.

\noindent
{\bf Requirement 1- Using uninflated planetesimals:} 
For similar integration times, the masses of the runaway growth bodies in simulations with uninflated 
planetesimals are considerably smaller than those with inflated planetesimals. For instance, as demonstrated in sections 3 and 4, 
the largest body reaches a mass of 142 times that of the initial planetesimals whereas in simulations
where planetesimals were initially inflated, this value varies between 300 and 1500. In the latter simulations,
the perturbation of these highly exaggerated bodies affect the dynamics of their surrounding in an unrealistic way.
Because when using inflated planetesimals, growth happens in much shorter time, these unreal
perturbations affect the evolution of the disk from the beginning, causing the formation, growth,
and orbital evolution of other bodies to be unrealistically perturbed as well. Using real-size planetesimals prevents all
this by allowing the system to evolve naturally and not be forced through an unrealistic dynamical path.

\noindent
{\bf Requirement 2- Including the effect of growing giant planets:}
In systems with 
uninflated planetesimals, the formation of large planetary embryos and the onset of the oligarchic growth 
takes several hundred thousands of years. For instance, in the simulations presented here, after 500,000 years of integrations,
the runaway growth was still in progress and the largest body reached a mass of only 142 times the mass of the 
initial planetesimals. In contrast, in simulations with inflated planetesimals 
\citep[e.g.,][]{Kokubo96,Wallace19}, bodies 3.5 to 120 times larger than ours formed during an integration time
that was 25 times shorter.
That means, to produce systems that would
show transition into the oligarchic growth, we had to continue the integrations much longer than 500,000 years. 
Similar long times for the growth of embryos were also reported by \citet{Kokubo02} where,
using an analytical model, the authors show that the growth timescale of embryos at 1 AU is of the order of 1 Myr and it
may reach 10 Myr at 5 AU for different radial profile of the disk surface density function.
As shown by \citet{Haghighipour12}, such integration 
times are comparable to the time that a growing giant planet in the orbit of Jupiter and Saturn manifests itself 
by perturbing the dynamical architecture of the asteroid belt. The latter strongly implies that in realistic simulations
of planetesimal growth, in addition to using uninflated planetesimals, integrations need to be carried out for long times
and include the perturbation of growing giant planets as well.

\noindent
{\bf Requirement 3- Simulating the entire disk at the same time:}
During an integration, the spatial distribution of planetesimals expands. 
In the systems of set A, where the initial distribution had a localized width of 0.08 AU, expansion was as large as 0.04 AU. 
In set B, initial distribution was more concentrated and had a width of 0.04 AU. In these systems, the expansion was about 0.06 AU. 
This spatial expansion reduces the surface density of planetesimals which causes their growth to slow down or even be hindered.
In a real disk, where the distribution of planetesimals expands over a much larger area, the drop in the local number/surface 
density of planetesimals is complemented by the expansion of adjacent local distributions and, as a result, the
overall surface density of the disk remains unchanged. This means, while simulating planetesimal growth in a localized
distribution can be helpful in unraveling the underlying physics of the processes, and while some remedies such as the
interpolation scheme presented by \citet{Clement20} can help with proofs of concepts, in real systems, simulations need to be 
carried out for the entire disk at the same time. 

\noindent
{\bf Requirement 4- Including dynamical friction due to the population of small bodies:}
In general, the eccentricities and inclinations of all large bodies stay low due to the dynamical friction with the 
remaining small planetesimals. In our simulations, the largest bodies maintained an eccentricity smaller than 
0.01 \citep[as a point of comparison, in the simulations of][this values ranged between 0.001 and 0.005]{Kokubo96,Kokubo00}. 
This is an important result that has significant implications for the formation of the final planets and their orbital architecture. 
It indicates that, although during an integration, the orbits of individual planetesimals are affected by the perturbation 
of the growing bodies, and although their numbers may drop by 40-60\%, still at any time during an integration, their population 
remains large enough for dynamical friction to manifest itself and be effective. In other words, small bodies such as small
planetesimals \citep[and in the modern simulations of planet formation, debris and small fragments created by collisions, see e.g.,][]{Crespi21} 
are essential parts of planet formation without which simulations will not produce reliable results.

\noindent
{\bf Requirement 5- Including nebular gas drag:}
Although, in order to maintain focus on the underlying physics of runaway growth, we did not consider the drag force 
of the nebula, we would like to note that any realistic simulation of planetesimal growth needs to include the effect
of the gas drag as well. Some authors have state that the inclusion of gas drag would only moderately
change the results \citep[e.g.,][]{Wallace19}. However, simulations presented here demonstrate that when this effect 
is not included, results will be noticeably different. For instance, the three largest bodies in the simulations of Section 3
reached the masses of 119-142 times the initial planetesimals in 500,000 years whereas in the simulations of 
\citet{Kokubo00}, where the authors used real-size planetesimals and gas drag was included, the largest body reached 
200 times the initial planetesimals in less than half the time of the integrations presented here($\sim 200,000$ years). 
This shorter time of the appearance of large bodies strongly implies that the subsequent dynamical evolution of the protoplanetary 
disk in systems with gas drag will also be considerably different from those that do not include this effect. 
We refer the reader to the Appendix for more details on the significance of this effect.

In conclusion, our analyses demonstrate that {\it accurate simulations of terrestrial planet formation need to 
start from a disk of planetesimals extended over a large area(e.g., from 0.5 au to at least 4.5 au),
use uninflated bodies (i.e., bodies with their actual sizes), and a mass-distributions that includes objects of different
sizes and masses. These simulations need to be carried out simultaneously with the growth of giant planets so that the
the gravitational perturbation of these bodies are taken into account. They also need to resolve collisions accurately,
by using, for instance, SPH (Smoothed Particle Hydrodynamics) simulations as in the works of \citet{Burger20} and
\citet{Reinhardt22}, include dynamical friction due to the background small bodies and the impact debris, and include the 
nebular gas drag as these two effects will contribute to the growth of planetesimals by damping their eccentricities and, 
therefore, their impact velocities.}

\vskip 10pt
\noindent
{\bf Acknowledgments.} We would like to express our deepest gratitude to Dr. Patryk Sofia Lykawka for his in-depth review of 
our manuscript and excellent recommendations that have greatly improved our paper. We are also thankful to the La Plata Institute 
for Astrophysics (IALP) for extensive use of their computational facilities and to the Information Technology division of the 
Institute for Astronomy at the University of Hawaii-Manoa for maintaining computational resources that were used in this project.
LAD acknowledges financial support from the Argentine National Institute for Advancement in Science and Technology
(ANPCyT) through grants PICT 2016-2635 and PICT 201-0505, the National University at La Plata, Argentina, 
(UNLP) through grant PID G144, and from The Department of Astronomical \& Geophysical Science at the UNLP
(FCAGLPUNLP). NH acknowledges support from NASA XRP through grant numbers 80NSSC18K0519, 80NSSC21K1050, and 80NSSC23K0270
and NSF grant AST-2109285.

\appendix
\section{The effect of nebular gas}

Because the growth of planetesimals begins at the early stage of the evolution of the protoplanetary disk,
due to their small sizes, the dynamics, and consequently, growth of these objects are strongly affected by their 
interaction with the nebula through gas drag. For a spherical body with a radius $R_{\rm p}$, 
the resistive force of the gas can be formulated as \citep{Landau59,Adachi76,Haghighipour03a}

\begin{equation}
{F_{\rm {drag}}}= -\,{1 \over 2}{C_{\rm D}}\pi{R_{\rm p}^2}\,{\rho_{\rm g}}({r_{\rm p}})\,{v_{\rm rel}^2}\,.
\end{equation}

\noindent
In this equation, $r_{\rm p}$ is the radial distance of the planetesimal to the central star, ${\rho_{\rm g}}({r_{\rm p}})$ 
is the density of the gas at the location of the planetesimal, and $v_{\rm rel}$ represents the velocity of the planetesimal
relative to the local gas velocity. The quantity $C_{\rm D}$ in equation (A1) is the drag coefficient whose
value depends on the size of the object and properties of the gas. When applied to the simulations of planetesimal
growth, it is customary to take ${C_{\rm D}}=2$. For more details on this quantity and its
values, we refer the reader to section 3.1 in \citet{Adachi76} and section 2.1 in \citet{Haghighipour03a}.

Studies of the effect of gas drag on the growth of planetesimals can be found in the fundamental works of 
\citet{Adachi76}, \citet{Ohtsuki93}, \citet{Kokubo00}, \citet{Thebault04} and \citet{Haghighipour05}. 
Collectively, these studies have demonstrated that gas drag has a positive effect on the rate of planetesimal
growth as it reduces their impact velocities by damping their orbital eccentricities and inclinations.
For instance, as shown by \citet{Thebault04}, gas drag aligns the periastrons of the orbits of planetesimals
reducing their mutual impact velocities, and thereby enhancing their collisional growth especially among objects
of the same size. The above works
have also shown that the damping effect of gas drag is smaller for larger objects meaning that at the
early stages of planetesimal growth, this effect has been more prominent.

The evidence to the fact that gas drag enhances collisional growth can be found in the results of the study by 
\citet{Kokubo00}. These authors presented the first numerical simulations in which uninflated planetesimals grow 
through perfect-merging while subject to the drag force of the nebula. Using the same initial set-up as those presented 
here (Section 2), these authors found that, after 200,000 years of integration, their largest runaway growth body reached the mass
of 200 times the initial planetesimals and after 500,000 years, this value rose up to several hundred. By comparison, 
in simulations of Section 3, the largest object is 142 times more massive than initial planetesimals and it took 470,000 years to form. 
Also, as shown by these authors, 
after 200,000 years of integration, the RMS values of the eccentricities and inclinations of planetesimals show 
a relatively flat distribution. In systems studied here, however, while after 250,000 years of integration, the RMS values of 
these quantities show, in general, a similar trend, their distribution is not flat and contains scattered values as well.

The comparison between the results presented here and those of \citet{Kokubo00} shows that while, as expected, results of the simulations
with and without gas drag are qualitatively similar, some subtle quantitative difference exist. These differences strongly
imply that in order to be able to build a realistic and quantitatively accurate model of terrestrial planet formation, the 
simulations of planetesimal growth need to include the effect of gas drag as well.

\clearpage
\begin{figure}
\vskip -45pt
\hskip -48pt
\includegraphics[scale=0.39]{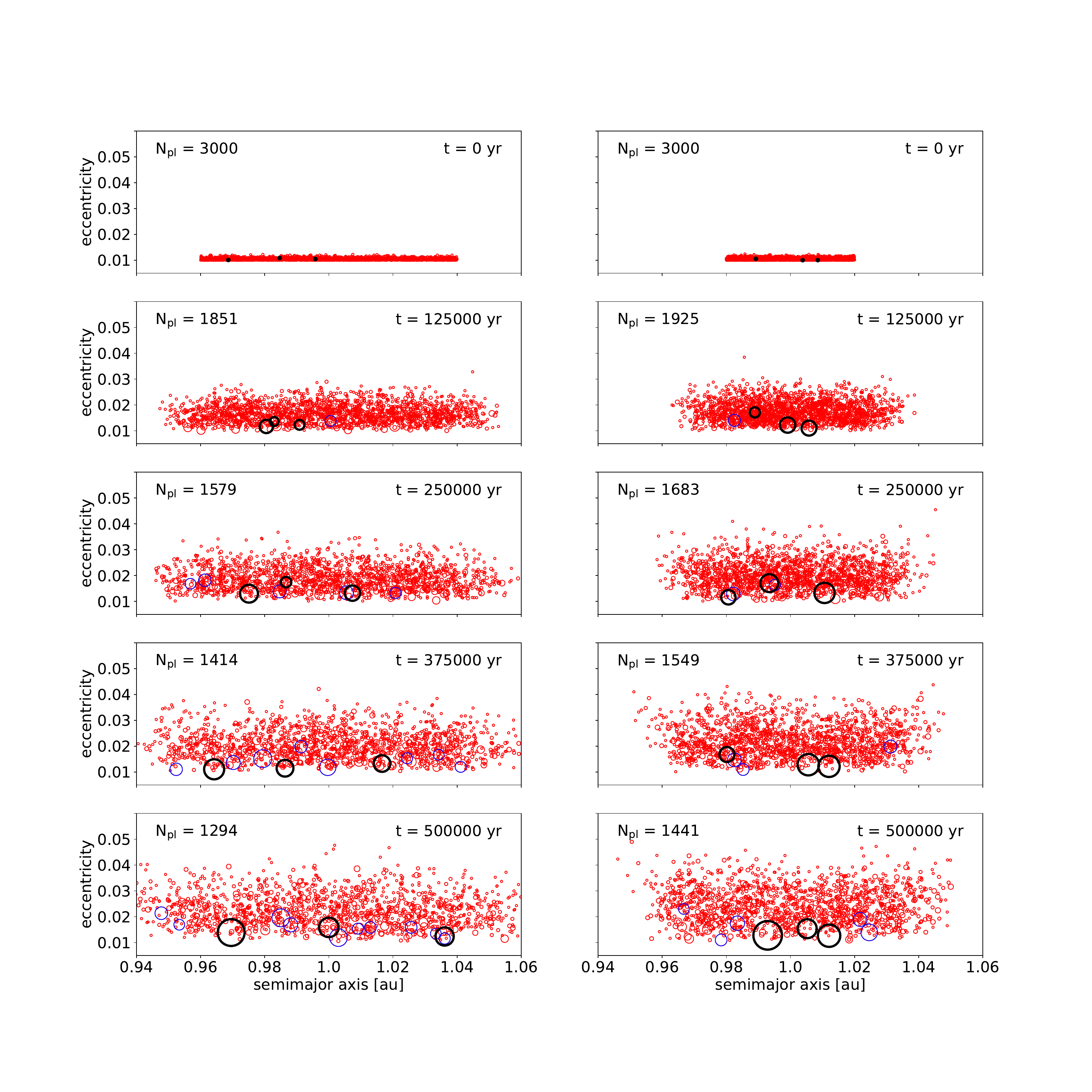}
\vskip -50pt
\caption{Snapshots of the evolution of systems A1 (left) and B4 (right) where the mass of the largest body is more 
than 100 times the initial planetesimals. Each object is represented by a red circle with its radius proportional to its mass. 
Blue circles represent bodies with masses at least 20 times their initial masses. The black circles in the bottom panel show 
the largest three bodies at the end of the integrations. Black circles in prior panels show the same bodies as they grow in 
time. Note the spreading of the disk and the dynamical diffusion of planetesimals (both red and blue circles) to outside the 
disk's initial boundaries during its evolution.}
\label{fig1}
\end{figure}

\clearpage
\begin{figure}
\vskip -50pt
\hskip -48pt
\includegraphics[scale=0.39]{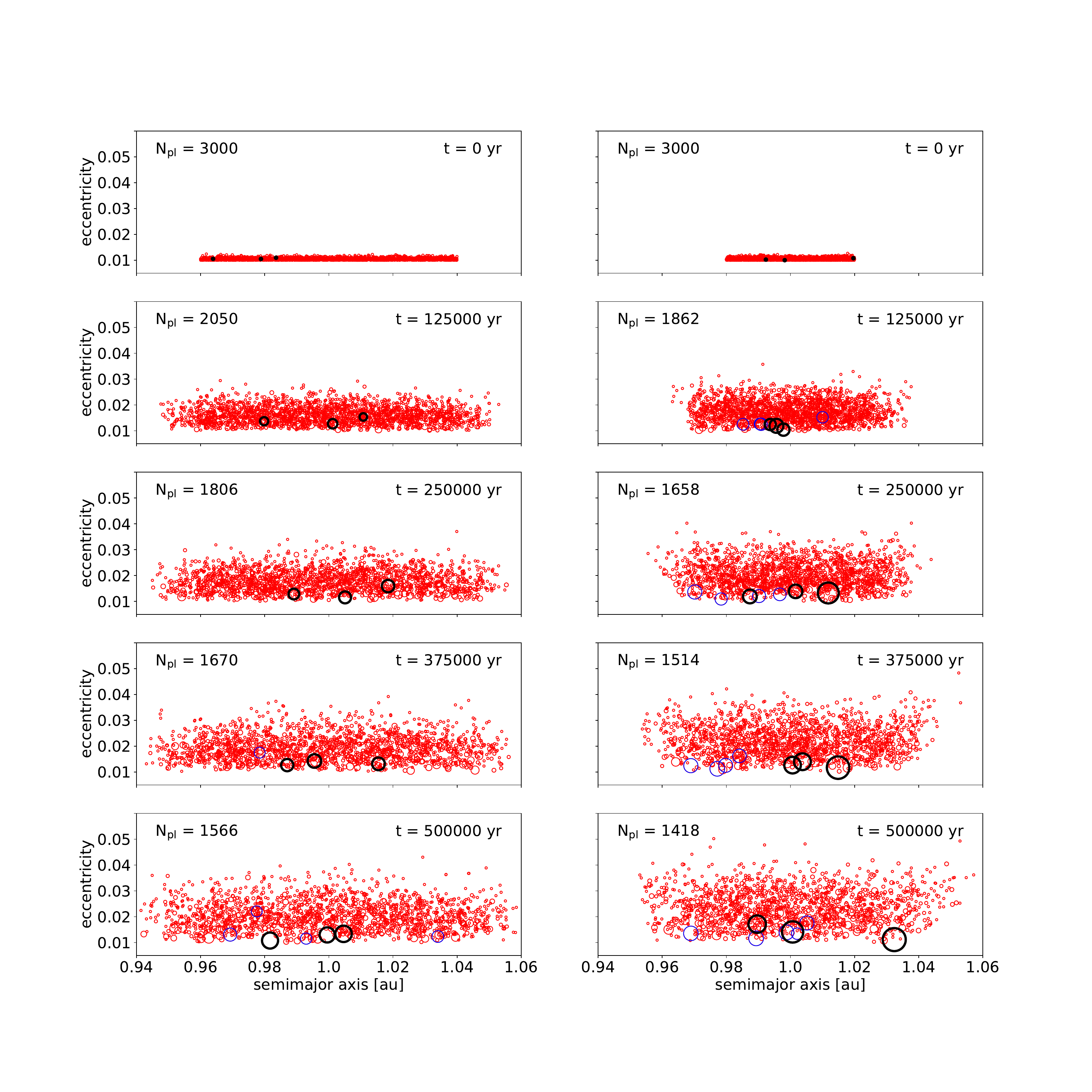}
\vskip -50pt
\caption{Same as figure 1, showing the evolution of systems A2 (left) and B5 (right) as samples of the rest of the systems.
Note the spreading of the disk and the dynamical diffusion of the planetesimals (red and blue circles) to outside the disk's
initial boundaries during its evolution.}
\label{fig2}
\end{figure}

\clearpage
\begin{figure}
\center
\includegraphics[scale=0.6]{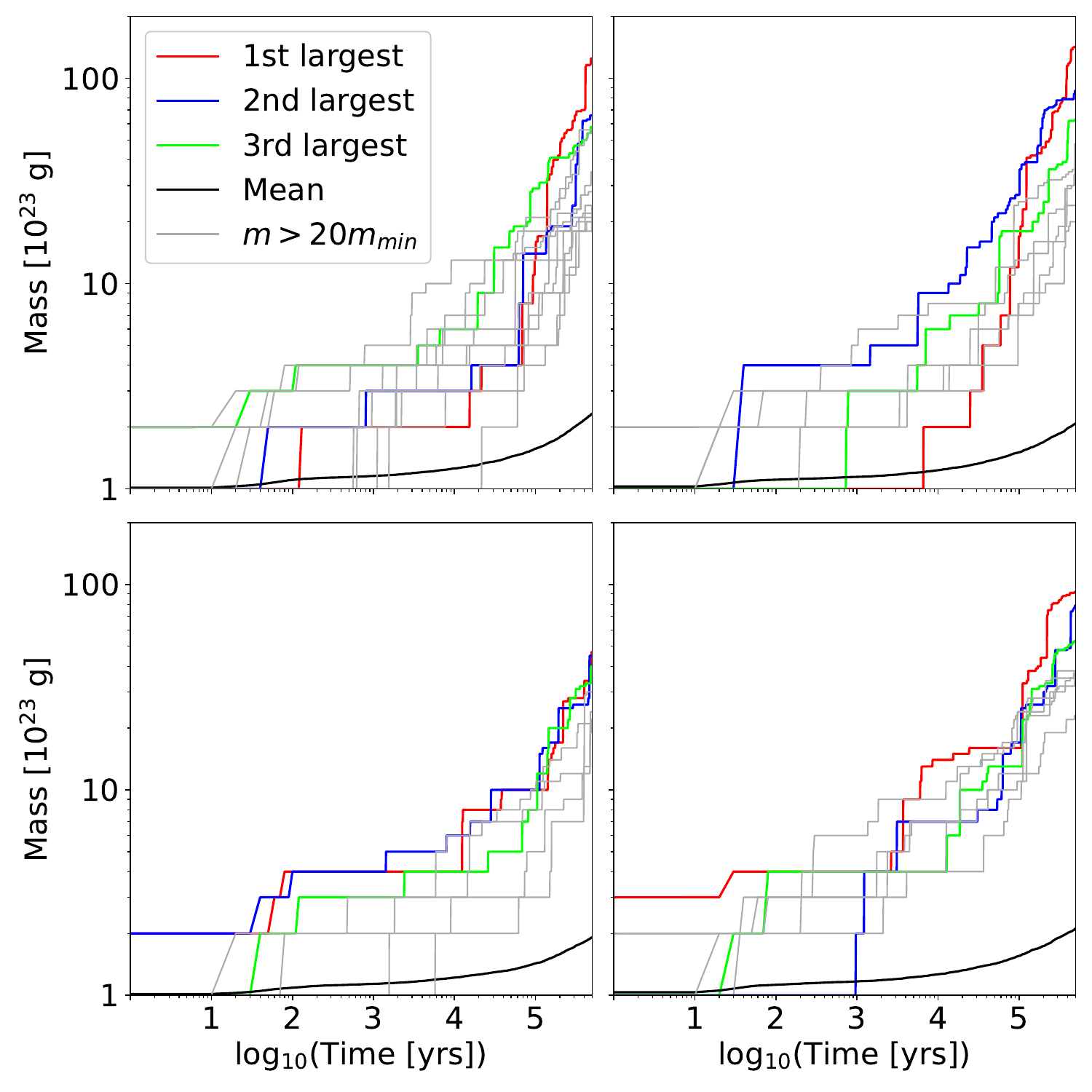}
\caption{Graphs of the growth of the large bodies in the systems of figures 1 and 2.
Top panels show systems A1 (left) and B4 (right), and bottom panels show systems A2 (left) and B5 (right).
The red, blue and green curves correspond to the three largest bodies (the black circles in figures 1 and 2)
and the curves in gray show the growth of bodies with a mass larger than 20 times the initial planetesimals $(m_{\rm min})$.
The black curve in each panel shows the variation of the mean mass of the system 
{\textit{without}} including the largest body.}
\label{fig3}
\end{figure}

\clearpage
\begin{figure}
\vskip -35pt
\hskip -123pt
\includegraphics[scale=0.9]{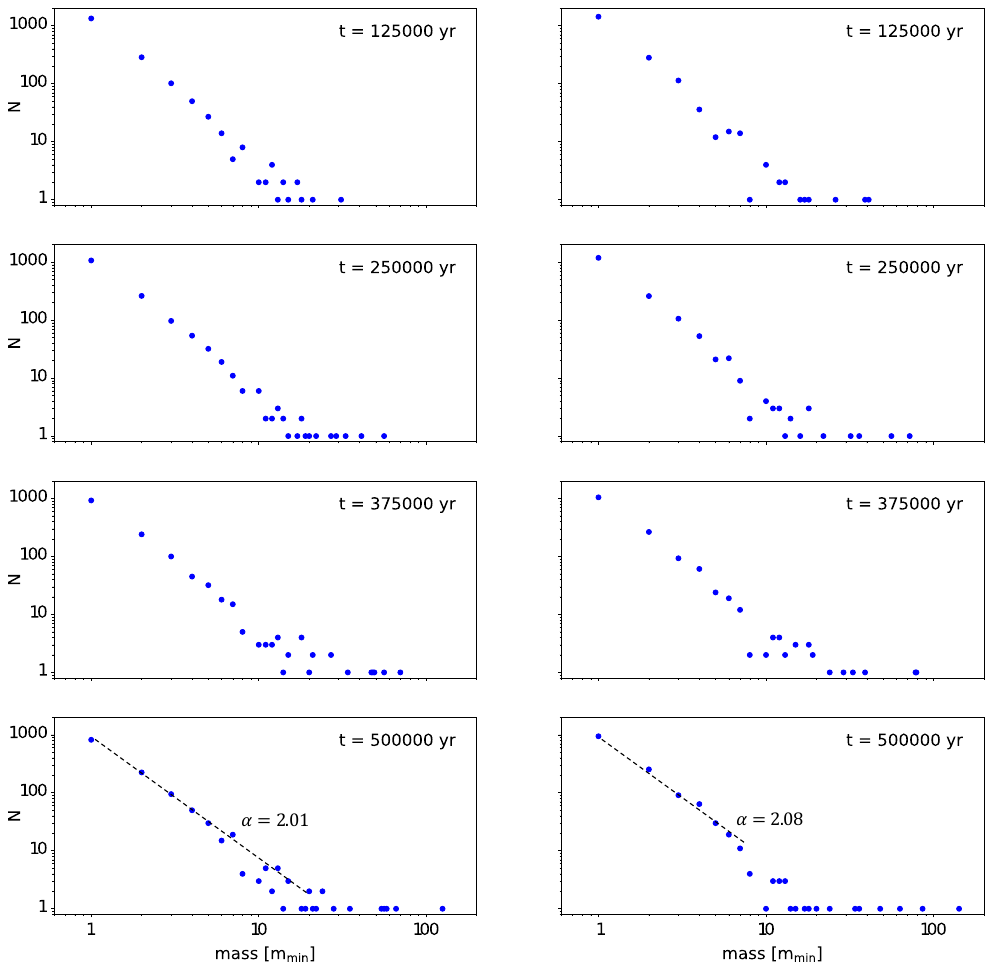}
\vskip -50pt
\caption{Graphs of the evolution of the mass distribution in systems A1 (left) and B4 (right).
Each point represents the number of bodies with that mass. Because growth is through perfect-merging, 
each mass is a multipe of the mass of the initial planetesimals  $(m_{\rm min})$. The bottom panels 
show the fits to the final mass distribution and their corresponding slopes (see Section 3.2 and Table 3
for more details).}
\label{fig4}
\end{figure}

\clearpage
\begin{figure}
\vskip -35pt
\hskip -123pt
\includegraphics[scale=0.9]{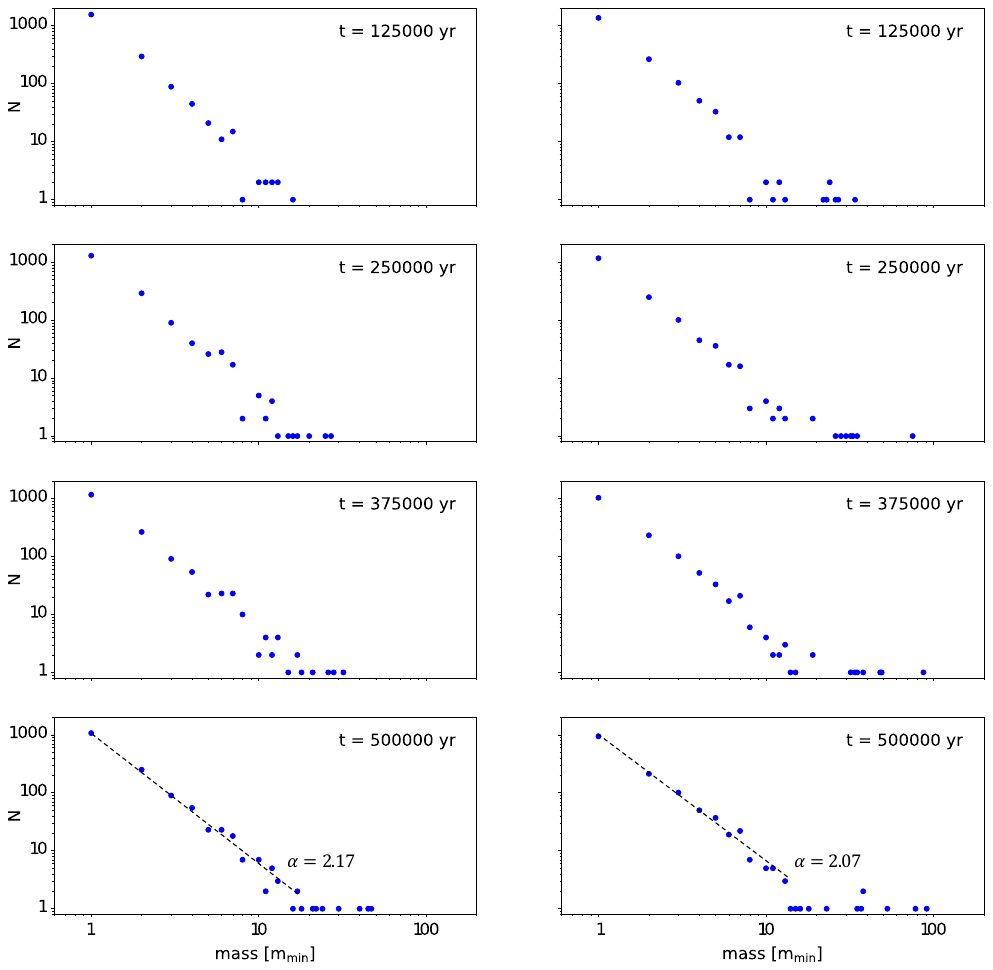}
\vskip -50pt
\caption{Graphs of the evolution of the mass distribution in systems A2 (left) and B5 (right).
Each point represents the number of bodies with that mass. Because growth is through perfect-merging, 
each mass is a multipe of the mass of the initial planetesimals  $(m_{\rm min})$. The bottom panels 
show the fits to the final mass distribution and their corresponding slopes (see Section 3.2 and Table 3
for more details).}
\label{fig5}
\end{figure}

\clearpage
\begin{figure}
\center
\includegraphics[scale=0.6]{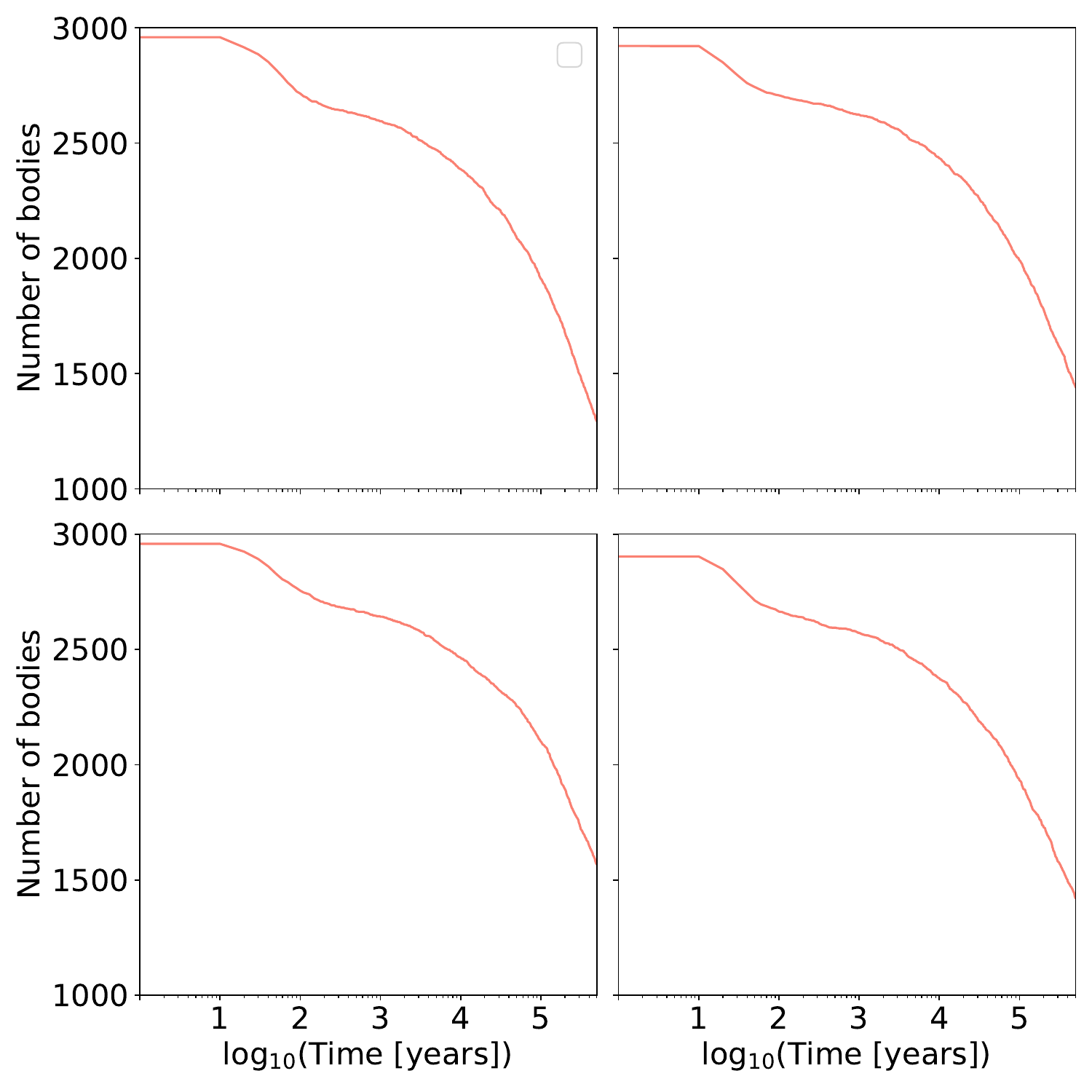}
\caption{Graphs of the number of the bodies in each system of figures 1 and 2 during the evolution of the system.
Top panels show systems A1 (left) and B4 (right), and bottom panels show systems A2 (left) and B5 (right).
Note that in systems A1 and B4 where the largest body is 100 times more massive than the initial planetesimals 
(as well as in simulation B5 where the largest body is 92 times more massive), 
the number of bodies drop by over 60\% whereas in simulation A2 where the largest body is 47 times the initial 
planetesimals, the number of bodies drop by only 50\%.}
\label{fig6}
\end{figure}

\clearpage
\begin{figure}
\vskip -55pt
\hskip -29pt
\includegraphics[scale=0.38]{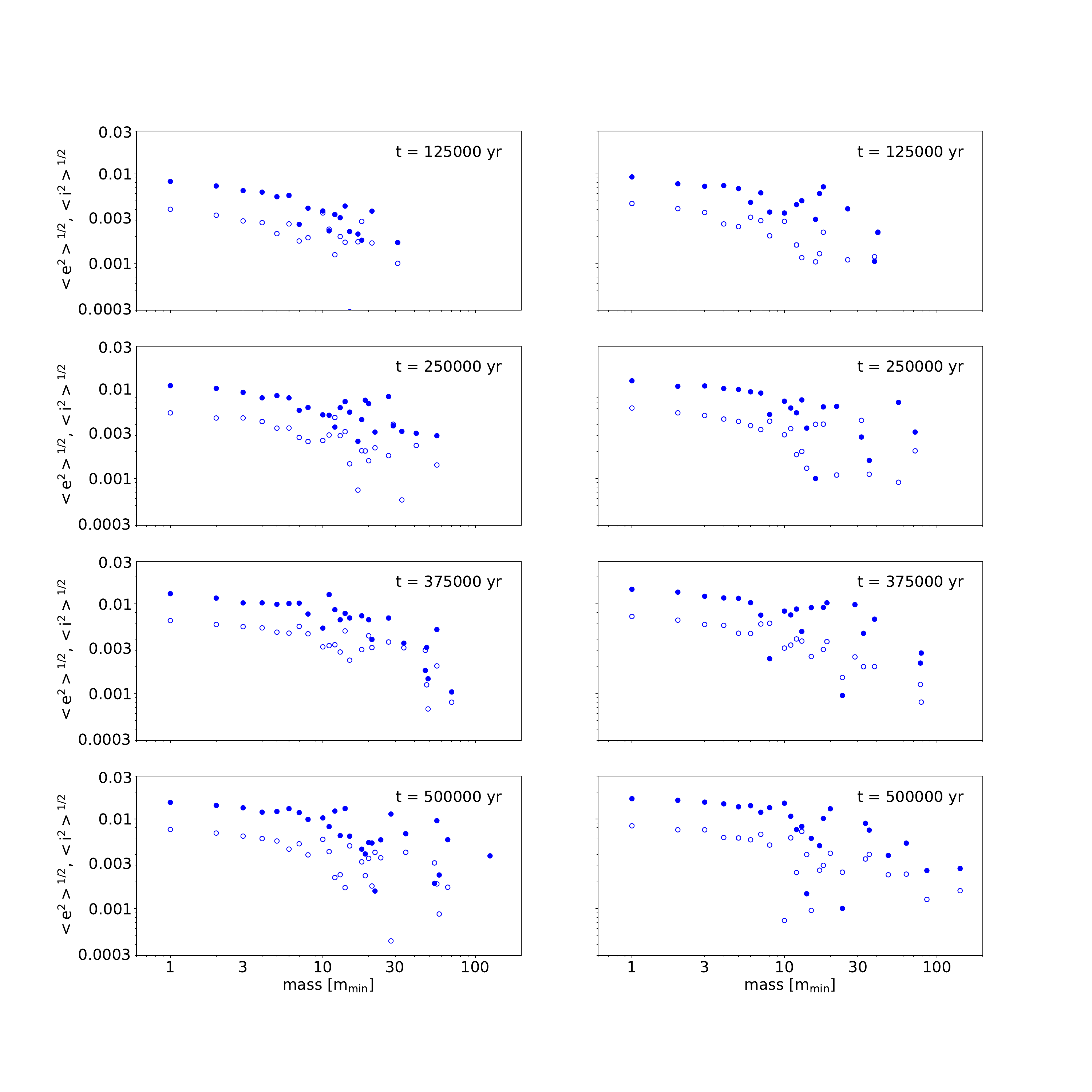}
\vskip -52pt
\caption{Graphs of the evolution of the RMS values of the eccentricities (filled circles) and inclinations (open circles)
of planetesimals in systems A1 (left) and B4 (right). Planetesimals masses have been binned as in figures 3 and 4. Note the 
damping of the eccentricity and inclination up to the point when the objects reach $10\, m_{\rm {min}}$. As shown by figures 4
and 5, this values of mass marks the onset of the runaway growth when the large bodies decouple from the rest of the planetesimals. 
At this stage, the perturbation of the larger bodies disturbs the dynamics of the planetesimals and their mutual
interactions cause their eccentricities and inclinations to increase. As the objetcs grows, their perturbing effects becomes 
stronger to the extent that the dynamical friction due to the remaining population of small planeteismals can hardly damp 
their eccentricities and inclinations to lower values.}
\label{fig7}
\end{figure}

\clearpage
\begin{figure}
\vskip -50pt
\hskip -29pt
\includegraphics[scale=0.38]{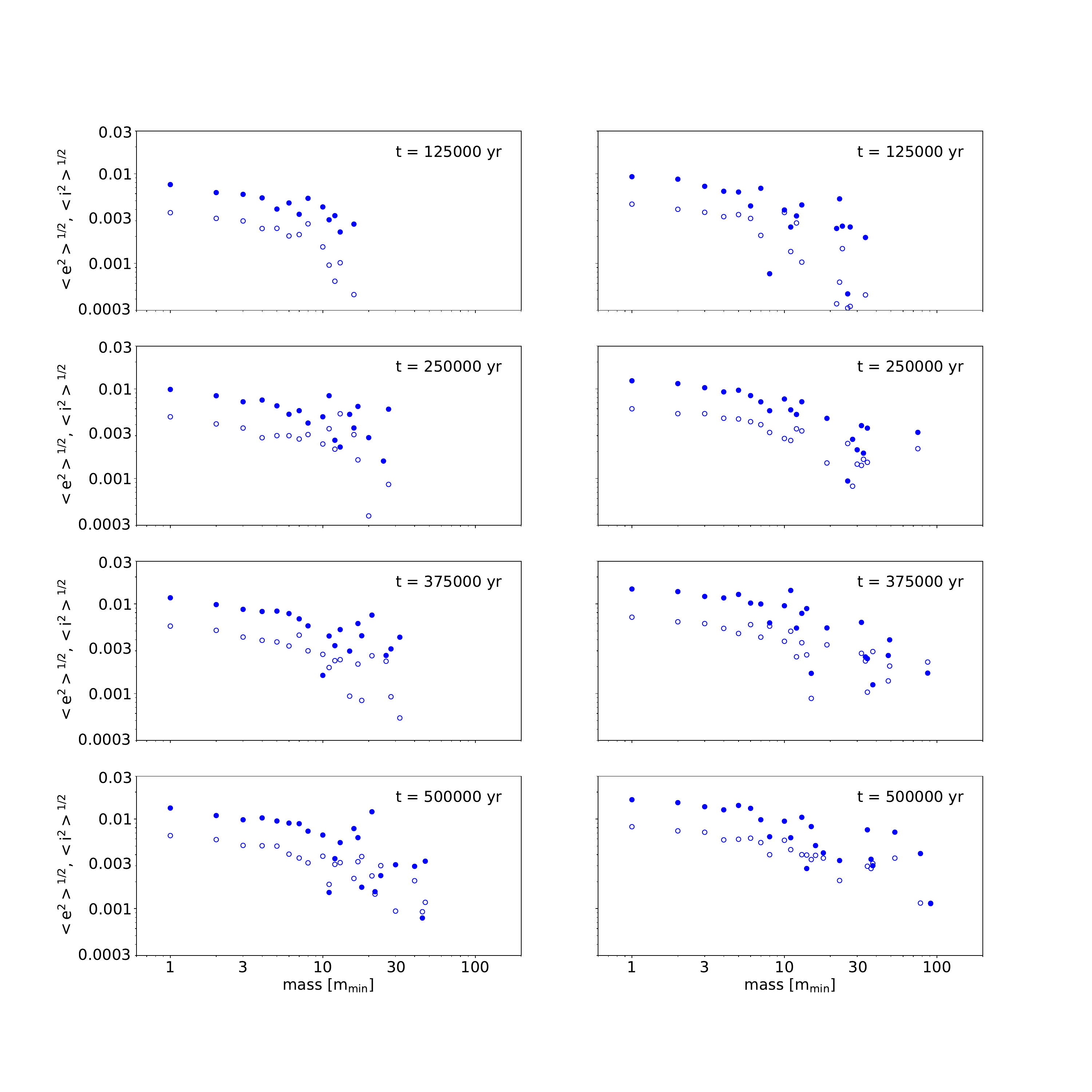}
\vskip -50pt
\caption{Graphs of the evolution of the RMS values of the eccentricities (filled circles) and inclinations (open circles)
of planetesimals in systems A2 (left) and B5 (right). Planetesimals masses have been binned as in figures 3 and 4. Note the 
damping of the eccentricity and inclination up to the point when the objects reach $10\, m_{\rm {min}}$. As shown by figures 4
and 5, this values of mass marks the onset of the runaway growth when the large bodies decouple from the rest of the planetesimals. 
At this stage, the perturbation of the larger bodies disturbs the dynamics of the planetesimals and their mutual
interactions cause their eccentricities and inclinations to increase. As the objetcs grows, their perturbing effects becomes 
stronger to the extent that the dynamical friction due to the remaining population of small planeteismals can hardly damp 
their eccentricities and inclinations to lower values.}
\label{fig8}
\end{figure}

\clearpage
\begin{deluxetable}{llll}
\tablenum{1}
\tablecaption{An up to date list of all $N$-body simulations of planetesimals growth showing the region of the 
simulation, integration resolution (number of planetesimals), and the radius-enlargement factor $(f)$.
\label{table1}}
\tablecolumns{4}
\tablewidth{0pt}
\tablehead{
\colhead{} & \colhead{Regions of} & \colhead{Number of} & \colhead{Enlargement} \\
\colhead{\hskip -115pt Reference} &  \colhead{Simulation} &  \colhead{Planetesimals} & \colhead {Factor} \\
\colhead{} &  \colhead{(AU)} & \colhead{} &  \colhead{$(f)$}}
\startdata
\citet{Kokubo96}     & \hskip 8pt $0.98-1.02$    & \hskip 28pt 3000   & \hskip 28pt 5       \\
\citet{Kokubo98}     & \hskip 8pt $0.978-1.02$   & \hskip 28pt 4000   & \hskip 28pt 4       \\
                     & \hskip 8pt $0.96-1.045$   & \hskip 28pt 4000   & \hskip 28pt 6       \\ 
\citet{Kokubo00}     & \hskip 8pt $0.99-1.01$    & \hskip 28pt 3000   & \hskip 28pt 1       \\
\citet{Richardson00} & \hskip 8pt $0.8-3.8$      & \hskip 28pt $10^6$ & \hskip 28pt 6        \\
\citet{Kokubo02}     & \hskip 8pt $0.96-1.04$    & \hskip 28pt $10^4$ & \hskip 28pt $6-10$   \\
\citet{Leinhardt05}  & \hskip 8pt $0-2.0$        & \hskip 28pt $10^4$ & \hskip 28pt 6        \\
                     & \hskip 8pt $0.915-1.085$  & \hskip 28pt 4000   & \hskip 28pt 6        \\
\citet{Morishima08}  & \hskip 8pt $0.7-1.3$      & \hskip 28pt 5000   & \hskip 28pt 4.3      \\
                     & \hskip 8pt $0.5-1.5$      & \hskip 28pt 5000   & \hskip 28pt 4.3      \\            
\citet{Barnes09}     & \hskip 8pt 0.4            & \hskip 28pt $10^5$ & \hskip 28pt 1        \\
\citet{Leinhardt09}  & \hskip 8pt $0-2.0$        & \hskip 28pt $10^4$ & \hskip 28pt 6        \\
\citet{Bonsor15}     & \hskip 8pt $0.5-1.5$      & \hskip 28pt ${10^4}-{10^5}$ & \hskip 28pt 6  \\
\citet{Carter15}     & \hskip 8pt $0.5-1.5$      & \hskip 28pt ${10^4}-{10^5}$ & \hskip 28pt 6  \\
\citet{Leinhardt15}  & \hskip 8pt $0.5-1.5$      & \hskip 28pt ${10^4}-{10^5}$ & \hskip 28pt 6  \\ 
\citet{Wallace19}    & \hskip 8pt $0.94-1.04$    & \hskip 28pt $10^6$ & \hskip 28pt 6    \\
\citet{Carter20}     & \hskip 8pt $0.5-3.0$      & \hskip 28pt $10^4$ & \hskip 28pt 1   \\
\citet{Clement20}    & \hskip 8pt 0.5, 1.0, 1.5, 2.0, 3.0 & \hskip 28pt $10^5$ & \hskip 28pt 1   \\
\enddata
\end{deluxetable}

\clearpage
\begin{deluxetable}{cccccc}
\scriptsize
\tablenum{2}
\tablecaption{The final mass and times of the growth of the three largest bodies (the black circles in figures 1 and 2) in all simulations.
The final mass is the mass of the body at the end of the 500,000 years integration and is given in terms of the planetesimals' initial mass
(${m_{\rm {min}}}= {10^{23}}$ g). Set A refers to initial distribution of planetesimals from 0.96 AU to 1.04 AU, and set B corresponds to an initial 
planetesimals distribution of 0.98 AU to 1.02 AU. 
\label{table2}}
\tablecolumns{6}
\tablewidth{0pt}
\tablehead{
\colhead{Set} & \colhead{Final Mass} & \colhead{Time of 25\%} & \colhead{Time of 50\%} &
\colhead{Time of 90\%} & \colhead{Time of Final Mass} \\
\colhead{} & \colhead{(Initial mass)} & \colhead{Growth (years)} & \colhead{Growth (years)} &
\colhead{Growth (years)} & \colhead{(years)}}
\startdata
   & 125 & 138840  & 286450  &  413540  &  481420 \\
A1 & 66  & 135070  & 313330  &  379090  &  471960 \\
   & 58  & 30780   & 91690   &  426950  &  478680 \\
\hline
   & 47  & 142150  & 218520  &  483980  &   490090 \\
A2 & 45  & 111790  & 191490  &  462390  &   462390 \\
   & 40  & 104750  & 144550  &  412480  &   481900 \\
\hline
   & 38  & 61810   & 183450  &  416020 &   492310 \\
A3 & 36  & 94000   & 134500  &  466990 &   488600 \\
   & 26  & 2330    & 33410   &  124680 &   284270 \\
\hline
   & 61  & 68360   & 152960  &  335650 &   460210 \\
A4 & 57  & 268990  & 334350  &  404290 &   477270 \\
   & 48  & 54229   & 214250  &  499770 &   499770 \\
\hline
   & 40  & 330510  & 330510  &  432240 &   498730 \\
A5 & 26  & 61840   & 266530  &  437260 &   468770 \\
   & 20  & 16610   & 95950   &  416480 &   416480 \\
\hline
   & 83  & 66390   &  239680 & 405040 &    491570 \\
B1 & 51  & 6740    &  64950  & 355930 &    469650 \\
   & 48  & 38050   &  214620 & 360980 &    457300 \\
\hline
   & 119 & 56070   & 302920 & 493750 &    494620 \\
B2 & 64  & 84770   & 353130  & 399390 &    490560 \\
   & 37  & 79970   & 185520  & 331280 &    411810 \\
\hline
   & 107 & 115880  & 132050 & 449090 &    492610 \\
B3 & 46  & 99469   & 245830  & 481370 &    481370 \\
   & 44  & 61620   & 110910  & 487790 &    487790 \\
\hline
   & 142 & 123370  & 303200 & 434940 &    473730 \\
B4 & 87  & 54780   & 168050 & 288670 &    490010 \\
   & 63  & 57230   &  232230 & 400370 &    487580 \\
\hline
   & 92  & 110540  & 221080 & 301400 &    481380 \\
B5 & 79  & 105860  & 279140 & 434760 &    488710 \\
   & 53  & 110280  & 135870 & 296740 &    475600 \\
\enddata
\end{deluxetable}

\clearpage
\begin{deluxetable}{ccc||ccc}
\tablenum{3}
\tablecaption{Values of $N_1$ and $\alpha$ for power-law fitting (equation 4) to the mass distributions 
at the end of all simulations. The values for $N_1$ were rounded to the nearest integer. 
\label{table3}}
\tablecolumns{6}
\tablewidth{0pt}
\tablehead{
\colhead{Sim} & \colhead{$N_1$} & \colhead{$\alpha$} & \colhead{Sim} & \colhead{$N_1$} & \colhead{$\alpha$}}
\startdata
 A1 & 1415 & 2.01 & B1 & 1489 & 2.14 \\
 A2 & 1618 & 2.17 & B2 & 1488 & 2.18 \\
 A3 & 1565 & 2.13 & B3 & 1460 & 2.12 \\
 A4 & 1567 & 2.17 & B4 & 1511 & 2.08 \\
 A5 & 1619 & 2.19 & B5 & 1445 & 2.07 \\
\enddata
\end{deluxetable}


\begin{thebibliography}{}

\bibitem[Aarseth et al. (1993)]{Aarseth93}
Aarseth SJ, Lin DNC and Palmer PL (1993) Evolution of Planetesimals. II. Numerical Simulations.
{\it Astrophysical Journal} {\bf 403}, 351-376.

\bibitem[Adachi et al. (1976)]{Adachi76}
Adachi I, Hayashi C and Nakazawa K (1976) The gas drag effect on the elliptical motion of a solid body in the primordial solar nebula. 
{\it Progress of Theoretical Physics} {\bf 56}, 1756-1771.

\bibitem[Barge and Pellat (1991)]{Barge91}
Barge B and Pellat R (1991) Mass spectrum and velocity dispersions during planetesimal accumulation I. Accretion. {\it Icarus}
{\bf 93}, 270-287.

\bibitem[Barnes et al. (2009)]{Barnes09}
Barnes R, Quinn T, Lissauer JJ and  Richardson D (2009) N-Body simulations of growth from 1 km planetesimals at 0.4 AU.
{\it Icarus} {\bf 203}, 626-643.

\bibitem[Batygin and Morbidelli (2023)]{Batygin23}
Batygin K and Morbidelli A (2023) Formation of rocky super-earths from a narrow ring of planetesimals 
{\it Nature Astronomy} {\bf 7}, 330 - 338

\bibitem[Beauge and Aarseth (1990)]{Beauge90}
Beauge C and Aarseth SJ (1990) N-Body Simulations of Planetary Formation. {\it Monthly Notices of the Royal Astronomical Society} 
{\bf 245}, 30-39.

\bibitem[Bonsor et al. (2015)]{Bonsor15}
Bonsor A, Leinhardt Z M, Carter P J, Elliott T, Walter M J and Stewart ST (2015) A collisional origin to Earth's non-chondritic composition?
{\it Icarus} {\bf 247}, 291-300.

\bibitem[Bromley and Kenyon (2006)]{Bromley06}
Bromley BC and Kenyon SJ (2006) A Hybrid N-Body-Coagulation Code for Planet Formation. {\it Astronomical Journal} {\bf 131}, 2737-2748.

\bibitem[Bromley and Kenyon (2011)]{Bromley11}
Bromley BC and Kenyon SJ (2011) A New Hybrid N-body-coagulation Code for the Formation of Gas Giant Planets. 
{\it Astrophysical Journal} {\bf 731}, 101.

\bibitem[Broz et al. (2022)]{Broz21}
Bro\u z M, Chrenko O, Nesvorny\'y D and Dauphas N (2021) Early terrestrial planet formation by torque-driven convergent migration of planetary 
embryos. {\it Nature Astronomy} {\bf 5}, 898–902

\bibitem[Burger et al. (2020)]{Burger20}
Burger C. Bazs\'o \'A and Sch\"afer CM (2020) Realistic collisional water transport during terrestrial planet formation. Self-consistent 
modeling by an N-body-SPH hybrid code. {\it Astronomy \& Astrophysics} {\bf 634}, A76. 

\bibitem[Carter et al. (2015)]{Carter15}
Carter PJ,  Leinhardt ZM,  Elliott T, Walter MJ and Stewart ST (2015) Compositional Evolution during Rocky Protoplanet Accretion.
{\it Astrophysical Journal} {\bf 813}, 72.

\bibitem[Carter and Stewart (2020)]{Carter20}
Carter PJ and Stewart ST (2020) Colliding in the Shadows of Giants: Planetesimal Collisions during the Growth and Migration of Gas Giants.
{\it Planetary Science Journal} {\bf 1}, 45.

\bibitem[Chambers (1999)]{Chambers99}
Chambers JE (1999) A hybrid symplectic integrator that permits close encounters between massive bodies.
{\it  Monthly Notices of the Royal Astronomical Society} {\bf 304}, 793-799.

\bibitem[Clement et al. (2020)]{Clement20}
Clement MS, Kaib NA and Chambers JE (2020) Embryo Formation with GPU Acceleration: Reevaluating the Initial Conditions for Terrestrial Accretion.
{\it Planetary Science Journal} {\bf 1}, 18.

\bibitem[Cox and Lewis (1980)]{Cox80}
Cox LP and Lewis J (1980) Numerical simulation of the final stages of terrestrial planet formation. {\it Icarus} {\bf 44}, 706-721.

\bibitem[Crespi et al. (2021)]{Crespi21}
Crespi S, Dobbs-Dixon I,  Georgakarakos N, Haghighipour N,  Maindl TI, Sch\"afer CM,  Winter PM (2021)
Protoplanet collisions: Statistical properties of ejecta. {\it Monthly Notices of the Royal Astronomical Society} {\bf 508}, 6013-6022.

\bibitem[Greenberg et al. (1978)]{Greenberg78}
Greenberg R, Wacker JF, Hartmann WK and Chapman CR (1978) Planetesimals to planets: Numerical simulation of collisional evolution. 
{\it Icarus} {\bf 35}, 1-26.

\bibitem[Haghighipour and  Boss (2003a)]{Haghighipour03a}
Haghighipour N and Boss AP (2003) On Pressure Gradients and Rapid Migration of Solids in a Nonuniform Solar Nebula.
{\it Astrophysical Journal} {\bf 583}, 996-1003.

\bibitem[Haghighipour and  Boss (2003b)]{Haghighipour03b}
Haghighipour N and Boss AP (2003) On Gas Drag Migration Solids in a Nonuniform Solar Nebula
{\it Astrophysical Journal} {\bf 598}, 1301-1311.

\bibitem[Haghighipour (2005)]{Haghighipour05}
Haghighipour N (2005) Growth and sedimentation of dust particles in the vicinity of local pressure enhancements in a solar nebula.
{\it Monthly Notices of the Royal Astronomical Society} {\bf 362}, 1015-1024.

\bibitem[Haghighipour and Scott (2012)]{Haghighipour12}
Haghighipour N and Scott ERD (2012) On the Effect of Giant Planets on the Scattering of Parent Bodies of Iron Meteorite from the Terrestrial 
Planet Region into the Asteroid Belt: A Concept Study. {\it Astrophysical Journal} {\bf 749}, 113.

\bibitem[Haghighipour and Winter (2016)]{Haghighipour16}
Haghighipour N and Winter OC (2016) Formation of terrestrial planets in disks with different surface density profiles.  
{\it Celestial Mechanics and Dynamical Astronomy} {\bf 124}, 235-268.

\bibitem[Haghighipour and Maindl (2022)]{Haghighipour22}
Haghighipour N and Maindl TJ (2022) Building Terrestrial Planets: Why Results of Perfect-merging Simulations Are Not Quantitatively Reliable 
Approximations to Accurate Modeling of Terrestrial Planet Formation. {\it Astrophysical Journal} {\bf 926}, 197.

\bibitem[Hayakawa et al. (1989)]{Hayakawa89}
Hayakawa M, Mizutani H, Kawakami S and Takagi Y (1989) Numerical simulation of collisional accretion process of the Earth. 
{\it Proceedings of the Lunar and Planetary Science Conference} {\bf 19}, 659.

\bibitem[Ida (1990)]{Ida90}
Ida S (1990) Stirring and dynamical friction rates of planetesimals in the solar gravitational field.
{\it Icarus} {\bf 88}, 129-145.

\bibitem[Ida and Makino (1992a)]{Ida92a}
Ida S and Makino J (1992a) N-Body simulation of gravitational interaction between planetesimals and a protoplanet . I. 
Velocity distribution of planetesimals. {\it Icarus} {\bf 96}, 107-120.

\bibitem[Ida and Makino (1992b)]{Ida92b}
Ida S and Makino J (1992b) N-body simulation of gravitational interaction between planetesimals and a protoplanet II. Dynamical friction.
{\it  Icarus} {\bf 98}, 28-37.

\bibitem[Ida and Makino (1993)]{Ida93}
Ida S Makino J (1993) Scattering of Planetesimals by a Protoplanet: Slowing Down of Runaway Growth.
{\it Icarus} {\bf 106}, 210-227.

\bibitem[Inaba et al. (1999)]{Inaba99}
Inaba S, Tanaka H, Ohtsuki K and Nakazawa K (1999) High-accuracy statistical simulation of planetary accretion: I. Test of the accuracy 
by comparison with the solution to the stochastic coagulation equation. {\it Earth, Planets and Space} {\bf 51}, 205-217.

\bibitem[Inaba et al. (2001)]{Inaba01}
Inaba S. Tanaka H. Nakazawa K. Wetherill GW and Kokubo E (2001) High-Accuracy Statistical Simulation of Planetary Accretion: II. 
Comparison with N-Body Simulation.  {\it Icarus} {\bf 149}, 235-250.

\bibitem[Izidoro et al. (2015)]{Izidoro15}
Izidoro A, Raymond SN, Morbidelli A and Winter OC (2015) Terrestrial planet formation constrained by Mars and the structure of the asteroid belt.
{\it Monthly Notices of the Royal Astronomical Society} {\bf 453}, 3619-3634.

\bibitem[Izidoro et al. (2022)]{Izidoro22}
Izidoro A, Dasgupta R, Raymond S N, Deienno R, Bitsch B and Isella A (2022) Planetesimal rings as the cause of the Solar System’s planetary 
architecture. {\it Nature Astronomy} {\bf 6}, 357–366.

\bibitem[Johansen et al. (2021)]{Johansen21}
Johansen A, Ronnet T, Bizzarro M, Schiller M, Lambrechts M, Nordlund \r A and Lammer, H (2021)
A pebble accretion model for the formation of the terrestrial planets in the Solar System. {\it Science Advances} {\bf 7}, 444. 

\bibitem[Kambara and Kokubo (2025)]{Kambara25}
Kambara y and Kokubo E (2025) Oligarchic growth of protoplanets in planetesimal rings {\it arXiv:2504.05667v1}

\bibitem[Kenyon and Luu (1998)]{Kenyon98}
Kenyon SJ and Luu JX (1998) Accretion in the Early Kuiper Belt. I. Coagulation and Velocity Evolution.
{\it Astronomical Journal} {\bf 115}, 2136-2160.

\bibitem[Kenyon and Luu (1999a)]{Kenyon99a}
Kenyon SJ and Luu JX (1999a) Accretion in the Early Kuiper Belt. II. Fragmentation. {\it Astronomical Journal} {\bf 118}, 1101-1119.

\bibitem[Kenyon and Luu (1999b)]{Kenyon99b}
Kenyon SJ and Luu JX (1998) Accretion in the Early Outer Solar System.  {\it Astrophysical Journal} {\bf 526}, 465-470.

\bibitem[Kenyon and Bromley (2001)]{Kenyon01}
Kenyon SJ and Bromley BC (2001) Gravitational Stirring in Planetary Debris Disks. {\it Astronomical Journal} {\bf 121}, 538-551.

\bibitem[Kenyon and Bromley (2002a)]{Kenyon02a}
Kenyon SJ and Bromley BC (2002a) Collisional Cascades in Planetesimal Disks. I. Stellar Flybys. {\it Astronomical Journal} {\bf 123}, 1757-1775.

\bibitem[Kenyon and Bromley (2002b)]{Kenyon02b}
Kenyon SJ and Bromley BC (2002b) Dusty Rings: Signposts of Recent Planet Formation. {\it Astrophysical Journal Letters} {\bf 577}, L35-L38.

\bibitem[Kenyon and Bromley (2004a)]{Kenyon04a}
Kenyon SJ and Bromley BC (2004a) Collisional Cascades in Planetesimal Disks. II. Embedded Planets. {\it Astronomical Journal} {\bf 127}, 513-530.

\bibitem[Kenyon and Bromley (2004b)]{Kenyon04b}
Kenyon SJ and Bromley BC (2004b) Detecting the Dusty Debris of Terrestrial Planet Formation. {\it Astrophysical Journal Letters} {\bf 602}, L133-L136.

\bibitem[Kenyon and Bromley (2006)]{Kenyon06}
Kenyon SJ and Bromley BC (2006) Terrestrial Planet Formation. I. The Transition from Oligarchic Growth to Chaotic Growth. 
{\it Astronomical Journal} {\bf 131}, 1837-1850.

\bibitem[Kobayashi et al. (2010)]{Kobayashi10}
Kobayashi H, Tanaka H, Krivov AV and Inaba S (2010) Planetary growth with collisional fragmentation and gas drag.
{\it Icarus} {\bf 209}, 836-847.

\bibitem[Kobayashi and L\"ohne (2014)]{Kobayashi14}
Kobayashi H and L\"ohne T (2014) Debris disc formation induced by planetary growth. 
{\it Monthly Notices of the Royal Astronomical Society} {\bf 442}, 3266-3274.

\bibitem[Kobayashi (2015)]{Kobayashi15}
Kobayashi H (2015) Orbital evolution of planetesimals in gaseous disks. 
{\it Earth, Planets and Space} {\bf 67}, 60.

\bibitem[Kobayashi et al. (2016)]{Kobayashi16}
Kobayashi H, Tanaka H and Okuzumi S (2016) From Planetesimals to Planets in Turbulent Protoplanetary Disks. I. Onset of Runaway Growth.
{\it Astrophysical Journal} {\bf 817}, 105.

\bibitem[Kobayashi and Tanaka (2018)]{Kobayashi18}
Kobayashi H and Tanaka H (2018) From Planetesimal to Planet in Turbulent Disks. II. Formation of Gas Giant Planets.
{\it Astrophysical Journal} {\bf 862}, 127.

\bibitem[Kokubo and Ida (1996)]{Kokubo96}
Kokubo E and Ida S (1996) On Runaway Growth of Planetesimals. {\it Icarus} {\bf 123}, 180-191.

\bibitem[Kokubo and Ida (1998)]{Kokubo98}
Kokubo E and Ida S (1998) Oligarchic Growth of Protoplanets. {\it Icarus} {\bf 131}, 171-178.

\bibitem[Kokubo and Ida (2000)]{Kokubo00}
Kokubo E and Ida S (2000) Formation of Protoplanets from Planetesimals in the Solar Nebula . {\it Icarus} {\bf 143}, 15-27.

\bibitem[Kokubo and Ida (2002)]{Kokubo02}
Kokubo E and Ida S (2002) Formation of Protoplanet Systems and Diversity of Planetary Systems. {\it Astrophysical Journal} {\bf 581}, 666-680.

\bibitem[Kokubo and Ida (2012)]{Kokubo12}
Kokubo E and Ida S (2012) Dynamics and accretion of planetesimals. {\it Progress of Theoretical and Experimental Physics}
{\bf 2012}, 01A308. 

\bibitem[Kolvoord and Greenberg (1992)]{Kolvoord92}
Kolvoord R A and Greenberg R (1992) A critical reanalysis of planetary accretion models. {\it Icarus} {\bf 98}, 2-19.

\bibitem[Kortenkamp et al. (2001)]{Kortenkamp01}
Kortenkamp SJ, Wetherill GW and Inaba S (2001) Runaway Growth of Planetary Embryos Facilitated by Massive Bodies in a Protoplanetary Disk.
{\it Science} {\bf 293}, 1127-1129.

\bibitem[Lambrechts et al. (2019)]{Lambrechts19}
Lambrechts M, Morbidelli A, Jacobson SA, Johansen A, Bitsch B, Izidoro A and Raymond SN (2019) 
Formation of planetary systems by pebble accretion and migration. How the radial pebble flux determines a terrestrial-planet 
or super-Earth growth mode. {\it Astronomy and Astrophysics} {\bf 627}, A83.

\bibitem[Landau and Lifshitz (1959)]{Landau59}
Landau LD and  Lifshitz EM (1959) {\it Fluid Mechanics}. London: Pergamon Press.

\bibitem[Lecar and Aarseth (1986)]{Lecar86}
Lecar M and Aarseth SJ (1986) A Numerical Simulation of the Formation of the Terrestrial Planets.
{\it Astrophysical Journal} {\bf 305}, 564-579.

\bibitem[Lee (2000)]{ManHoi00}
Lee M-H (2000) On the Validity of the Coagulation Equation and the Nature of Runaway Growth. {\it Icarus} 143, 74-86.

\bibitem[Leinhardt et al. (2000)]{Leinhardt00}
Leinhardt ZM, Richardson DC and Quinn T (2000) Direct N-body Simulations of Rubble Pile Collisions. 
{\it Icarus} {\bf 146}, 133-151.

\bibitem[Leinhardt and Richardson (2002)]{Leinhardt02}
Leinhardt ZM and Richardson DC (2002) N-Body Simulations of Planetesimal Evolution: Effect of Varying Impactor Mass Ratio.
 {\it Icarus} {\bf 159}, 306-313.

\bibitem[Leinhardt and Richardson (2005)]{Leinhardt05}
Leinhardt ZM and Richardson DC (2005) Planetesimals to Protoplanets. I. Effect of Fragmentation on Terrestrial Planet Formation.
{\it Astrophysical Journal} {\bf 625}, 427-440.

\bibitem[Leinhardt et al. (2009)]{Leinhardt09}
Leinhardt ZM, Richardson DC, Lufkin G and Haseltine J (2009) Planetesimals to protoplanets - II. 
Effect of debris on terrestrial planet formation. {\it Monthly Notices of the Royal Astronomical Society} {\bf 396}, 718-728.

\bibitem[Leinhardt and Stewart (2012)]{Leinhardt12}
Leinhardt ZM and Stewart ST (2012) Collisions between Gravity-dominated Bodies. I. Outcome Regimes and Scaling Laws.
{\bf 745}, 79.

\bibitem[Leinhardt et al. (2015)]{Leinhardt15}
Leinhardt ZM, Robinson J, Carter PJ and Lines S (2015) Numerically Predicted Indirect Signatures of Terrestrial 
Planet Formation. {\it Astrophysical Journal} {\bf 806}, 23.

\bibitem[Levison and Agnor (2003)]{Levison03}
Levison HF and Agnor C (2003) The Role of Giant Planets in Terrestrial Planet Formation.
{\it Astronomical Journal}, {\bf 125}, 2692-2713.

\bibitem[Levison et al. (2015a)]{Levison15a}
Levison HF, Kretke KA and Duncan MJ (2015a) Growing the gas-giant planets by the gradual accumulation of pebbles.
{\it Nature} {\bf 524}, 322-324.

\bibitem[Levison et al. (2015b)]{Levison15b}
Levison HF, Kretke KA, Walsh KJ and Bottke WF (2015b) Growing the terrestrial planets from the gradual 
accumulation of sub-meter sized objects. {\it Proceedings of the National Academy of Sciences} {\bf 112}, 14180-14185.

\bibitem[Matsumura et al. (2017)]{Matsumura17}
Matsumura S, Brasser R and Ida S (2017) N-body simulations of planet formation via pebble accretion. I. First results.
{\it Astronomy and Astrophysics} {\bf 607}, A67.

\bibitem[Morbidelli et al. (2018)]{Morbidelli18}
Morbidelli A, Nesvorny D, Laurenz V, Marchi S, Rubie DC, Elkins-Tanton L, Wieczorek M and Jacobson S
(2018) The timeline of the lunar bombardment: Revisited. {\it Icarus} {\bf 305}, 262-276.

\bibitem[Morbidelli et al. (2022)]{Morbidelli22}
Morbidelli A, Bailli\'e B, Batygin K, Charnoz S, Guillot T, Rubie D C and Kleine T (2022) Contemporary formation of early 
Solar System planetesimals at two distinct radial locations. {\it Nature Astronomy} {\bf 6}, 72–79

\bibitem[Morbidelli et al. (2025)]{Morbidelli25}
Morbidelli A, Kleine T and Nimmo E (2025) Did the terrestrial planets of the solar system form by pebble accretion?
{\it Earth and Planetary Science Letters} {\bf 650}, 119120

\bibitem[Morishima et al. (2008)]{Morishima08}
Morishima R, Schmidt MW, Stadel J and Moore B (2008) Formation and Accretion History of Terrestrial Planets 
from Runaway Growth through to Late Time: Implications for Orbital Eccentricity. {\it Astrophysical Journal} {\bf 685}, 1247-1261.

\bibitem[Morishima et al. (2010)]{Morishima10}
Morishima R, Stadel, J and Moore B (2010) From planetesimals to terrestrial planets: $N$-body simulations including the effects 
of nebular gas and giant planets. {\it Icarus} {\bf 207}, 517-535.

\bibitem[Nakagawa et al. (1983)]{Nakagawa83}
Nakagawa Y, Hayashi C and Nakazawa K (1983) Accumulation of planetesimals in the solar nebula. {\it Icarus} {\bf 54}, 361-376.

\bibitem[Ogihara et al. (2018)]{Ogihara18}
Ogihara1 M, Kokubo E, Suzuki T K and Morbidelli A (2018) Formation of the terrestrial planets in the solar system around
1 au via radial concentration of planetesimals. {\it Astronomy \& Astrophysics} {\bf 612}, L5 

\bibitem[Ohtsuki et al. (1988)]{Ohtsuki88}
Ohtsuki K, Nakagawa Y and Nakazawa K (1988) Growth of the earth in nebular gas. {\it Icarus} {\bf 75}, 552-565.

\bibitem[Ohtsuki and Ida (1990)]{Ohtsuki90}
Ohtsuki K and Ida S (1990) Runaway planetary growth with collision rate in the solar gravitational field.
{\it Icarus} {\bf 85}, 499-511.

\bibitem[Ohtsuki et al. (1993)]{Ohtsuki93}
Ohtsuki K, Ida S, Nakagawa Y, Nakazawa K (1993) Planetary Accretion in Solar Gravitational Field. 
in: {\it Protostars and Planets III} Eds: E. H. Levy and J. I. Lunine, Tucson, AZ. University of Arizona Press, PP.1089 

\bibitem[Ohtsuki et al. (2002)]{Ohtsuki02}
Ohtsuki K, Stewart GR and Ida S (2002) Evolution of Planetesimal Velocities Based on Three-Body Orbital Integrations
and Growth of Protoplanets. {\it Icarus} {\bf 155}, 436-453.

\bibitem[Ormel et al. (2010a)]{Ormel10a}
Ormel CW, Dullemond CP and Spaans M (2010a) Accretion among preplanetary bodies: The many faces of runaway growth.
 {\it Icarus} {\bf 210}, 507-538.

\bibitem[Ormel et al. (2010b)]{Ormel10b}
Ormel CW, Dullemond CP Spaans M (2010b) A New Condition for the Transition from Runaway to Oligarchic Growth.
{\it Astrophysical Journal Letters} {\bf 714}, L103-L107.

\bibitem[Ormel and Okuzumi (2013)]{Ormel13}
Ormel CW and Okuzumi S (2013) The Fate of Planetesimals in Turbulent Disks with Dead Zones. II. Limits on the 
Viability of Runaway Accretion.  {\it Astrophysical Journal} {\bf 77}, 44.

\bibitem[Palmer et al. (1993)]{Palmer93}
Palmer PL, Lin DNC, Aarseth SJ (1993) Evolution of Planetesimals. I. Dynamics: Relaxation in a Thin Disk. 
{\it Astrophysical Journal} {\bf 403}, 336-350.

\bibitem[Rafikov (2003a)]{Rafikov03a}
Rafikov RR (2003) The Growth of Planetary Embryos: Orderly, Runaway, or Oligarchic? 
{\it Astronomical Journal} {\bf 125}, 942-961.

\bibitem[Rafikov (2003b)]{Rafikov03b}
Rafikov RR (2003) Dynamical Evolution of Planetesimals in Protoplanetary Disks. 
{\it Astronomical Journal} {\bf 126}, 2529-2548.

\bibitem[Reinhardt et al. (2022)]{Reinhardt22}
Reinhardt C, Meier T, Stadel JG, Otegi JF and Helled, R (2022) Forming iron-rich planets with giant impacts.
{\it Monthly Notices of the Royal Astronomical Society} {\bf 517}, 3132-3143.

\bibitem[Richardson et al. (2000)]{Richardson00}
Richardson DC, Quinn T, Stadel J and Lake G (2000) Direct Large-Scale N-Body Simulations of Planetesimal Dynamics. 
{\it Icarus} {\bf 143}, 45-59.

\bibitem[Safronov (1962)]{Safronov62}
Safronov VS (1962) {\it A particular Solution of the Coagulation Equations}. {\it Doklady Akademii nauk SSSR.} {\bf 147}, 64

\bibitem[Safronov (1969)]{Safronov69}
Safronov VS (1969) {\it Evolution of Protoplanetary Cloud and Formation of the Earth and the Planets}.
Nauka, Moscow

\bibitem[Spaute et al. (1991)]{Spaute91}
Spaute D, Weidenschilling SJ, Davis DR and Marzari F (1991) Accretional evolution of a planetesimal swarm: 
1. A new simulation  {\it Icarus} {\bf 92}, 147-164.

\bibitem[Stewart and Kaula (1980)]{Stewart80}
Stewart GR and Kaula WM (1980) A gravitational kinetic theory for planetesimals. {\it Icarus} {\bf 44}, 154-171.

\bibitem[Stewart and Leinhardt (2012)]{Stewart12}
Stewart ST and Leinhardt ZM (2012) Collisions between Gravity-dominated Bodies. II. The Diversity of Impact 
Outcomes during the End Stage of Planet Formation.  {\it Astrophysical Journal} {\bf 751}, 32.

\bibitem[Tanaka and Nakazawa (1994)]{Tanaka94}
Tanaka H and Nakazawa K (1994) Validity of the Statistical Coagulation Equation and Runaway Growth of Protoplanets.
{\it Icarus} {\bf 107}, 404-412.

\bibitem[Thebault et al. (2004)]{Thebault04}
Th\'ebault P, Marzari F, Scholl H, Turrini D and Barbieri M (2004) Planetary formation in the $\gamma$ Cephei system. 
{\it Astronomy and Astrophysics} {\bf 427}, 1097-1104.

\bibitem[Wallace and Quinn (2019)]{Wallace19}
Wallace SC and Quinn TR (2019) N-body simulations of terrestrial planet growth with resonant dynamical friction. 
{\it Monthly Notices of the Royal Astronomical Society} {\bf 489}, 2159-2176.

\bibitem[Walsh and Levison (2019)]{Walsh19}
Walsh KJ and Levison HF (2019) Planetesimals to terrestrial planets: Collisional evolution amidst a dissipating gas disk.
{\it Icarus} {\bf 329}, 88-100.

\bibitem[Weidenschilling et al. (1997)]{Weidenschilling97}
Weidenschilling SJ, Spaute D, Davis DR, Marzari F and Ohtsuki K (1997) Accretional Evolution of a Planetesimal Swarm. 
{\it Icarus} {\bf 128}, 429-455.

\bibitem[Wetherill (1990)]{Wetherill90}
Wetherill GW (1990) Formation of the earth. {\it Annual Review of Earth and Planetary Sciences} {\bf 18}, 205-256.

\bibitem[Wetherill and Stewart (1989)]{Wetherill89}
Wetherill GW and Stewart GR (1989) Accumulation of a swarm of small planetesimals. {\it Icarus} {\bf 77}, 330-357.

\bibitem[Wetherill and Stewart (1993)]{Wetherill93}
Wetherill GW and Stewart GR (1993) Formation of Planetary Embryos: Effects of Fragmentation, Low Relative Velocity,
and Independent Variation of Eccentricity and Inclination  {\it Icarus} {\bf 106}, 190-209

\end{thebibliography}
\end{document}